\documentclass[12pt]{spieman}  % 12pt font required by SPIE;
\usepackage{amsmath,amsfonts,amssymb}
\usepackage{graphicx}
\usepackage{setspace}
\usepackage{tocloft}

\newcommand{\nustar}{\textit{NuSTAR}}

\title{Reconstruction of the NuSTAR point spread function using single-laser metrology}

\author[a]{Hannah P. Earnshaw}
\author[a,b]{Kristin K. Madsen}
\author[a]{Karl Forster}
\author[a]{Brian W. Grefenstette}
\author[a]{Murray Brightman}
\author[c]{Andreas Zoglauer}
\author[a]{Fiona Harrison}
\affil[a]{Cahill Center for Astronomy and Astrophysics, California Institute of Technology, Pasadena, CA 91125, USA}
\affil[b]{Goddard Space Flight Center, Greenbelt, MD 20771, USA}
\affil[c]{Space Sciences Laboratory, University of California, Berkeley, CA 94720, USA}

\cftpagenumbersoff{figure}
\cftpagenumbersoff{table} 
\begin{document} 
\maketitle

\begin{abstract}

This paper describes a method by which the metrology system of the {\it Nuclear Spectroscopic Telescope Array} (\nustar) X-ray space observatory, which uses two lasers to characterize the relative motion of the optics and focal plane benches, can be approximated should one laser fail. The two benches are separated by a ten-meter-long rigid mast that undergoes small amounts of thermal flexing which need to be compensated for in order to produce a non-blurred image. We analyze the trends of mast motion by archival observation parameters in order to discover whether the mast motion in future observations can be predicted. We find that, by using the solar aspect angle (SAA), observation date, and orbital phase, we can simulate the motion of one laser by translating the track produced by the other and applying modifications to the resulting mast aspect solution, allowing the reconstruction of a minimally distorted point spread function in most cases. We will implement the generation of simulated mast files alongside the usual \nustar\ data reduction pipeline for contingency purposes. This work has implications for reducing the risk of implementing laser metrology systems on future missions that use deployable masts to achieve the long focal lengths required in high-energy astronomy by mitigating the impact of a metrology laser failure in the extended phase of a mission.

\end{abstract}

% Include a list of up to six keywords after the abstract

% Include email contact information for corresponding author
{\noindent \footnotesize\textbf{*}Hannah Earnshaw,  \linkable{hpearn@caltech.edu} }

\begin{spacing}{2}   % use double spacing for rest of manuscript

\section{Introduction}
\label{sec:intro}  

The Nuclear Spectroscopic Telescope Array (\nustar) is a NASA Small Explorer mission\cite{harrison13} and the first space-based telescope to be able to focus hard X-rays (3--79\,keV). The long focal length required to make this possible is achieved by the telescope being made up of two benches, one containing the focusing optics (henceforth, the optics bench) and the other containing the detectors at the focal plane (henceforth, the focal plane bench, with detectors FPMA and FPMB), separated by a 10.15\,m rigid mast which was deployed once the spacecraft was in orbit. 

While the carbon fiber mast is designed for minimal thermal flexing, there is still a small amount of motion over time as the spacecraft orbits the Earth, moves into and out of sunlight, and points at targets across the sky at a wide range of angles relative to the Sun. If not accounted for, this motion would cause images taken by \nustar\ to be blurred as the positions of the optics and focal plane benches change relative to each other. In order to successfully reconstruct an in-focus image, we need to know the relative positions and orientations of the two benches at all times. This is done using a laser metrology system\cite{liebe10,harp10}, made up of two laser-detector pairs that measure the mast position several times a second over the course of an observation.

\begin{figure}
\begin{center}
\includegraphics[width=14cm]{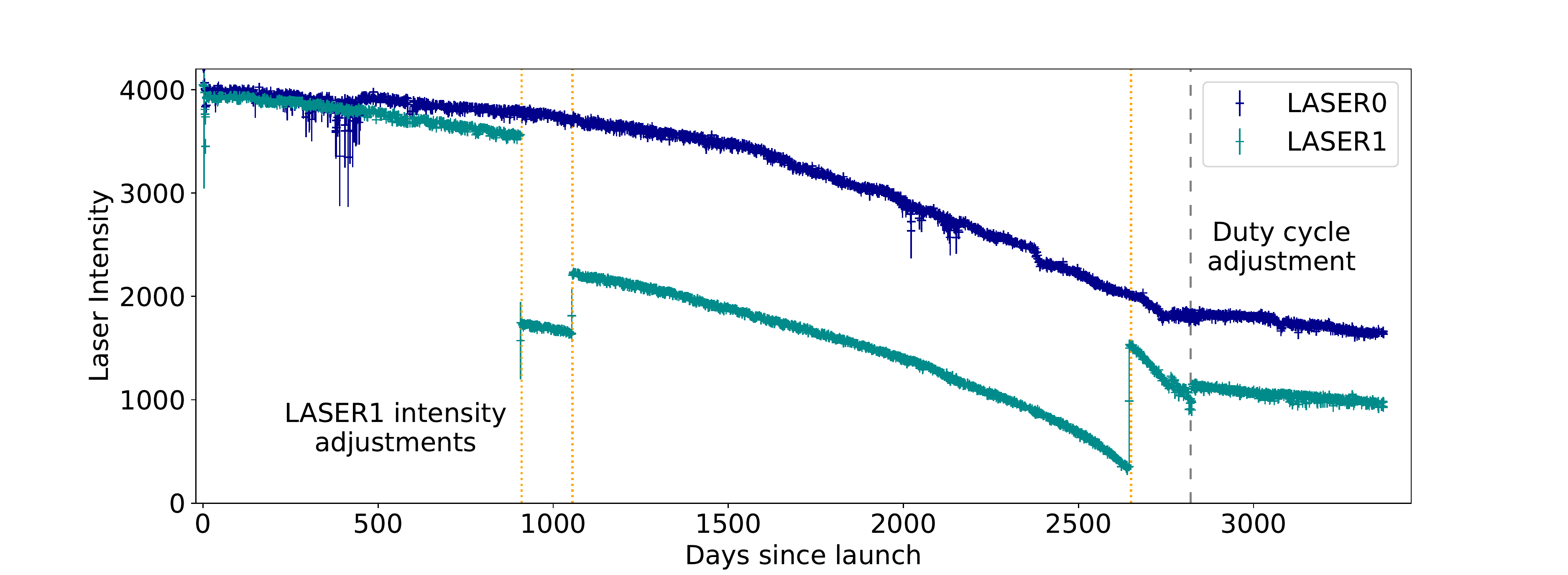}
\end{center}
\caption 
{ \label{fig:lasers} The laser intensity over the duration of the mission for LASER0 (blue) and LASER1 (cyan) as measured by the respective position sensitive detectors (see Section~\ref{sec:metrology}). Also indicated are times that the current to LASER1 was adjusted (dotted orange) and when the laser duty cycle was adjusted (dashed grey). At the present time, the intensity continues to decline, albeit slower than previously. } 
\end{figure}

The laser metrology system for \nustar\ has worked well over the mission's lifetime. However, the intensity of both lasers as measured at their respective position-sensitive detectors (PSDs) has declined since the beginning of the mission (Fig.~\ref{fig:lasers}), with LASER0 operating at a little under one half of its original intensity and LASER1 at approximately a quarter. While steps have successfully been taken to slow this decline, such as altering the duty cycle of the laser cycling, the cause is not fully understood. Although the risk of a metrology laser failing before the end of the mission lifetime is low, as a contingency for laser failure we want to be able to reconstruct a source point spread function (PSF) using only the data from a single metrology laser, in addition to any other knowledge of the telescope's condition we have at our disposal.

\begin{figure}
\begin{center}
\includegraphics[width=12cm]{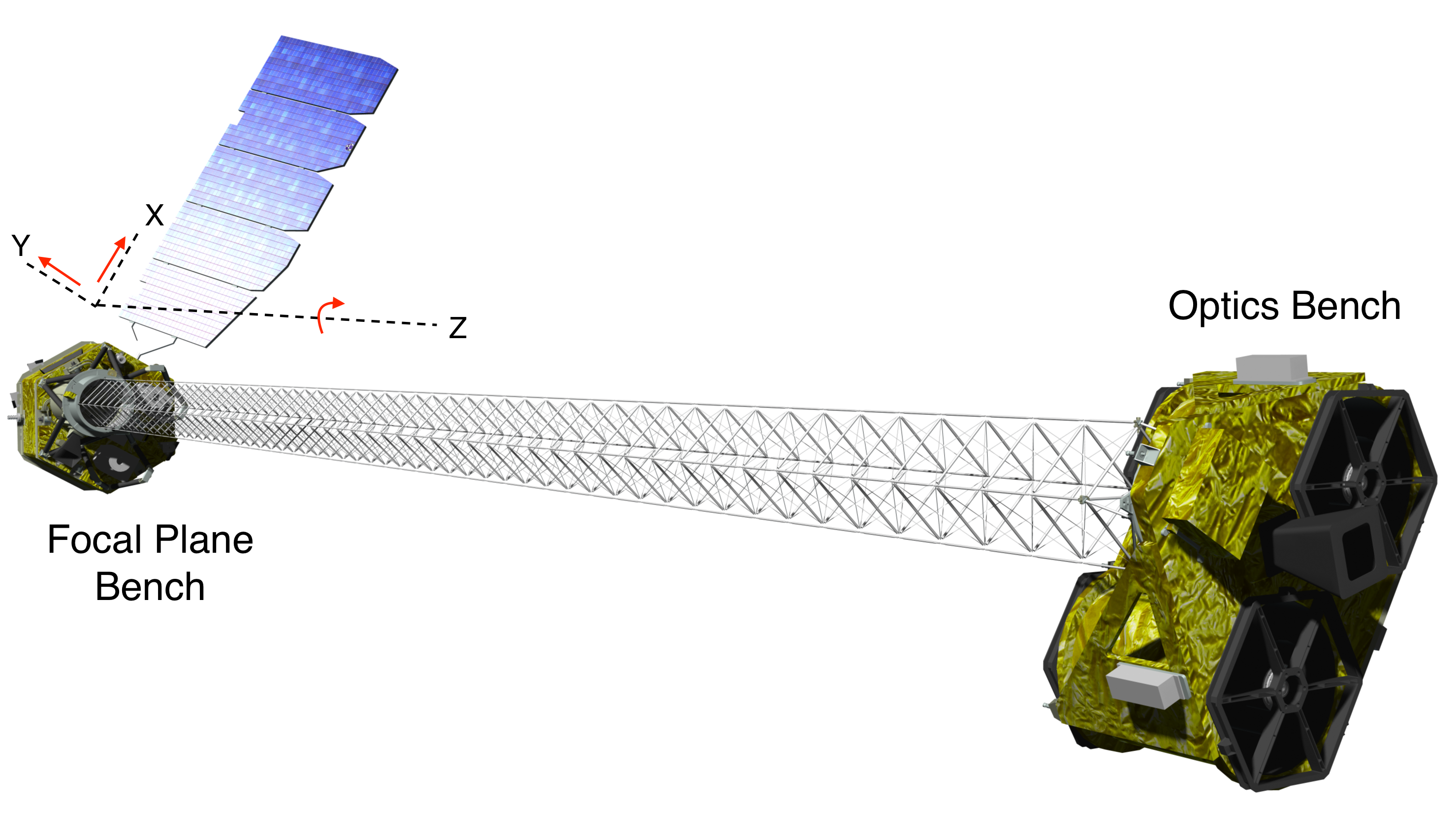}
\end{center}
\caption 
{ \label{fig:nustar}
An illustration of the \nustar\ telescope, showing its two benches and mast. The X, Y and Z axes we use to refer to motion between the benches are indicated with black dashed lines, and the three degrees of freedom that are measured by the metrology system are show with red arrows.} 
\end{figure}

In this paper, we describe the metrology system in detail in Section~\ref{sec:metrology}.  We then present a method of reconstructing the \nustar\ PSF using a single metrology laser, the solar aspect angle (SAA), and the mission elapsed time in Section~\ref{sec:method}, and demonstrate its effectiveness in Section~\ref{sec:results}. We discuss the consequences of which laser is used for reconstruction in Section~\ref{sec:reverse}. Finally, we summarize and discuss potential future implementation in Section~\ref{sec:summary}. For all examples shown, we use an observation of GX~13+1, taken in 2017 (Obsid: 30301003002). 

\section{Metrology System}
\label{sec:metrology}  

Between the optics bench and the focal plane bench of \nustar, there are six degrees of freedom of motion that could conceivably be measured, but only three that are significant enough to affect the image reconstruction: translation in the X and Y axes (that is, the axes orthogonal to the mast) and the rotation around the Z axis (that is, the degree to which the mast twists) \cite{liebe10}. We illustrate the telescope, its axes, and the degrees of freedom we measure in Fig.~\ref{fig:nustar}. The mast length and angle from parallel of the two benches are sufficiently constant and deviations small that the effect on the aspect reconstruction is negligible.

\begin{figure}
\begin{center}
\includegraphics[width=18cm]{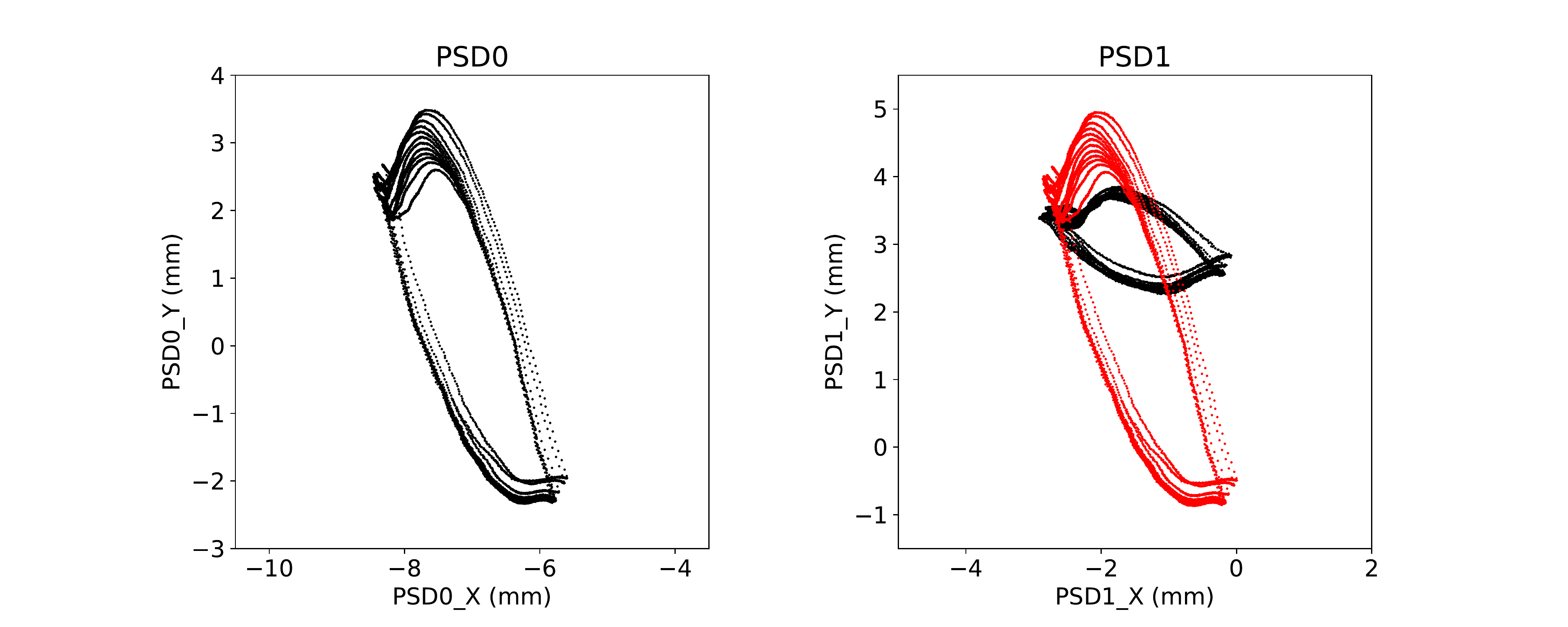}\vspace{-4mm}
\end{center}
\caption 
{\label{fig:psdtracks}
The tracks of the two laser spots on their PSDs for observation 30301003002. The original tracks are shown in black, and the track from PSD0 overlaid at the position of the PSD1 track is shown in red. The tracks have been thinned to show 10\% of the data points, for clarity.} 
\end{figure}

These degrees of freedom are measured using two lasers mounted on the optics bench (henceforth LASER0 and LASER1), shining on two independent PSDs mounted on the focal place bench (PSD0 and PSD1 respectively). The PSDs are 20mm photodiodes that can measure the centroid of a light spot on their surface in two dimensions \cite{liebe12}. The position of the centroid of the laser spot is measured at a frequency of 4\,Hz and recorded, after being corrected for the response of the PSDs, in the Corrected PSD File ({\tt psdcorr}) as part of the data reduction pipeline using the \nustar\ Data Analysis Software (NuSTARDAS). Example laser spot tracks are shown in black in Fig.~\ref{fig:psdtracks}.

These tracks are used to calculate the mast aspect solution that relates a photon vector at the focal plane bench $\vec{v}_{\rm fb}$ to a photon vector at the optical bench $\vec{v}_{\rm ob}$ \cite{harp10,forster16}. The mast solution is described by a translation $T_{\rm mast} = [T_x, T_y, T_z]$ (with $T_z$ fixed to the mast length), and a quaternion $Q_{\rm mast} = [q0, q1, q2, q3]$ where $q3$ is the real component and thus the mast twist angle $\theta = 2\cos^{-1}(q3)$, such that:

\begin{equation}
\vec{v}_{\rm fb} = Q_{\rm mast} \vec{v}_{\rm ob} + \vec{T}_{\rm mast}
\end{equation}

Aspect reconstruction is therefore the inverse of this process in order to recreate the image made by photons entering the telescope at the optics bench (in addition to the consideration of the sky-to-optics-bench and focal-plane-bench-to-detector transforms, which we are not concerned with for the purposes of this study). Aspect reconstruction is performed for each individual photon based on the values of $Q_{\rm mast}$ and $T_{\rm mast}$ at the photon time of arrival. The values of $Q_{\rm mast}$ and $T_{\rm mast}$ over time are stored in the Mast Aspect Solution File ({\tt mast}) for each observation. 

\begin{figure}
\begin{center}
\includegraphics[width=16cm]{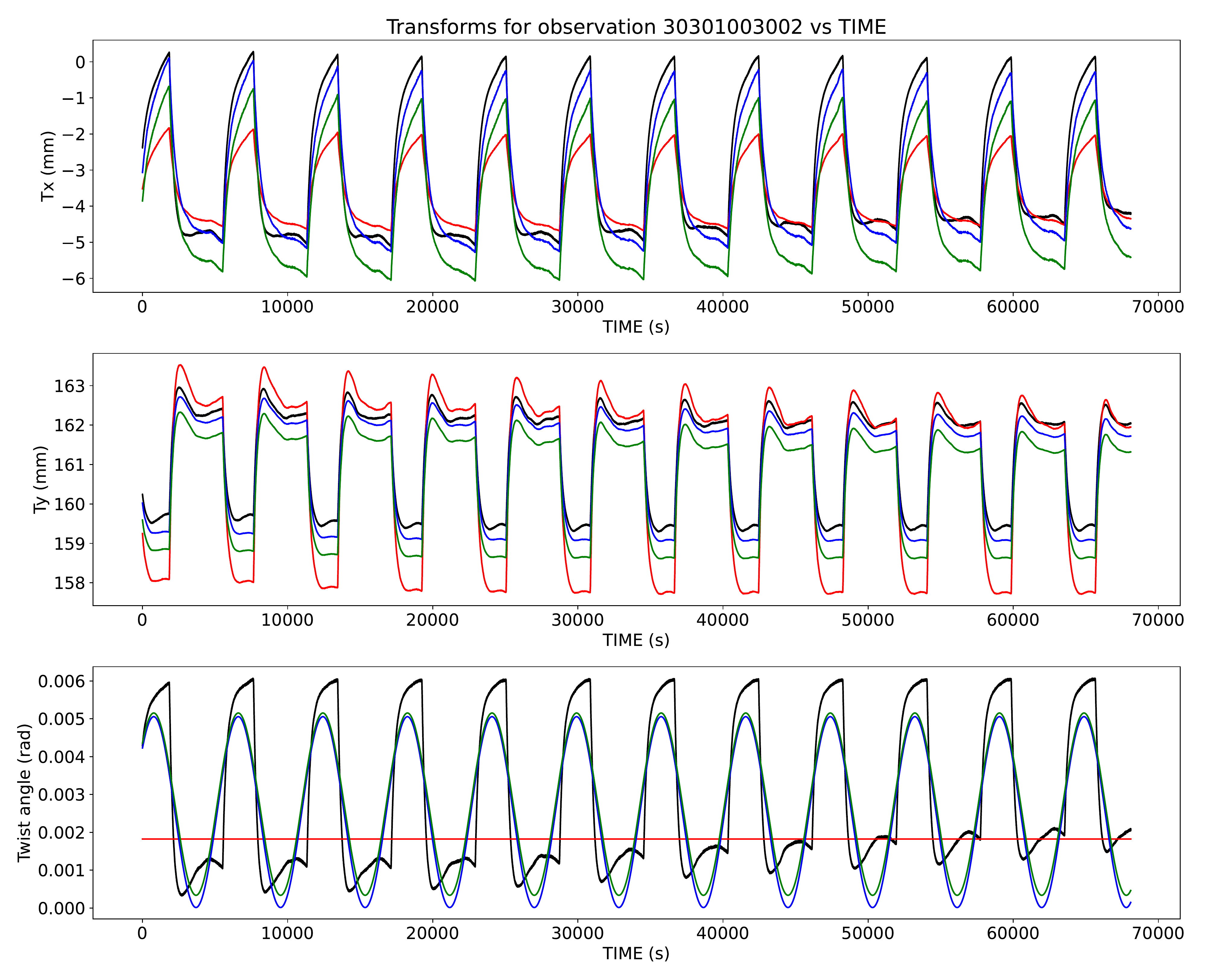}
\vspace{-4mm}
\end{center}
\caption 
{\label{fig:masttracks}
The values of the X transform $T_x$, Y transform $T_y$, and mast twist angle $\theta$ over time for observation 30301003002. The original two-laser solution is plotted in black, the mast values derived from the single-laser PSD tracks are plotted in red, the ideal corrected values are plotted in blue, and the corrected values estimated using the SAA are plotted in green.} 
\end{figure}

The mast motion is due to thermal flexing caused by changes in the illumination of the telescope by the Sun, both from occultation by the Earth each orbit and from self-shadowing, which changes as a function of Solar aspect angle (SAA). The SAA is the angle between the pointing of the spacecraft and the Earth-Sun axis (where SAA = 0 means that the spacecraft is pointing directly at the Sun, and SAA = 180 is directly anti-Sun) and remains constant during an observation. The roll angle does not contribute to thermal variations since the Sun is always kept on the same side of the spacecraft, normal to the Solar arrays, such that the direction of the Sun around the Z-axis does not change\cite{harrison13}. Therefore, the changes in $T_x$, $T_y$, and $\theta$ over the course of a single observation are both periodic with the orbital period and reasonably consistent in shape after settling due to the SAA remaining constant (see Fig.~\ref{fig:masttracks}, black). This means that the mast motion is complex but to an extent predictable, since the orbit and SAA are known for each observation. This predictability gives us the tools to compensate for the potential loss of a laser, and to successfully reconstruct an image with just a single metrology laser.

\section{Single-laser Reconstruction Method}
\label{sec:method}

The method we use to perform aspect reconstruction using a single metrology laser (e.g. LASER0) is as follows: we assume that LASER1 has failed and create a synthetic PSD1 track by translating the measured PSD0 track to the would-be position of the PSD1 track, as predicted from the date and SAA (see Section~\ref{sec:psdtranslation}). We use this track to generate a new {\tt psdcorr} file and use it as an input to the NuSTARDAS module {\tt numetrology} to generate a new {\tt mast} file. We make modifications to this new {\tt mast} file based on the SAA (see Section~\ref{sec:mast}) and use this corrected {\tt mast} file as input to {\tt nupipeline} which proceeds to run as it normally would to generate aspect corrected event lists. It is important to note that these corrections do not affect the energy reconstruction but only the photon distribution on the focal plane detector. 

\subsection{PSD Track Translation}
\label{sec:psdtranslation}

\begin{figure}
\begin{center}
\includegraphics[width=12cm]{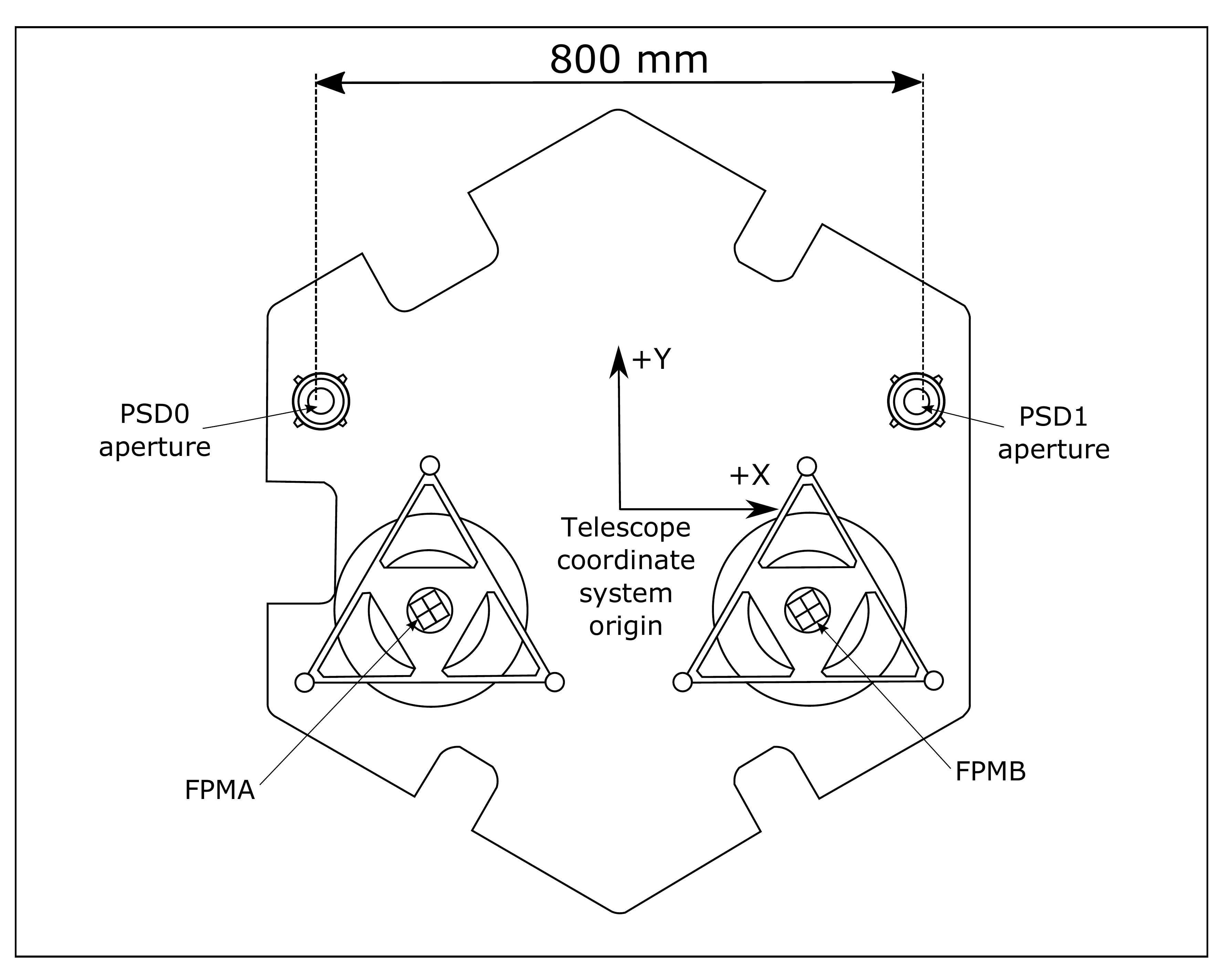}
\vspace{-4mm}
\end{center}
\caption 
{\label{fig:fpm}
The layout and selected components of the focal plane module of \nustar, showing the location of the FPM detectors and the PSDs, as well as the origin of the telescope coordinate system. Viewed from the positive Z direction (i.e. the direction of the optics bench).} 
\end{figure}

The first step in this process is to translate the PSD0 track to the position of the PSD1 track. This is done by using the quaternions and vectors stored in the appropriate CALDB alignment file for the observation date to transform the track from PSD0 coordinates to focal plane bench coordinates, apply a vector to move the PSD0 track to the position of the PSD1 track, and transform this track back to PSD1 coordinates (see Fig.~\ref{fig:psdtracks}). The position of the PSDs and the layout of the focal plane bench is shown in Fig.~\ref{fig:fpm}. In order to determine the length of the baseline separating the tracks and the angle of rotation from the X axis, we measured the median of both PSD tracks in focal bench coordinates for all observations taken over the course of the \nustar\ mission. 

\begin{figure}
\begin{center}
\vspace{-4mm}
\includegraphics[width=14cm]{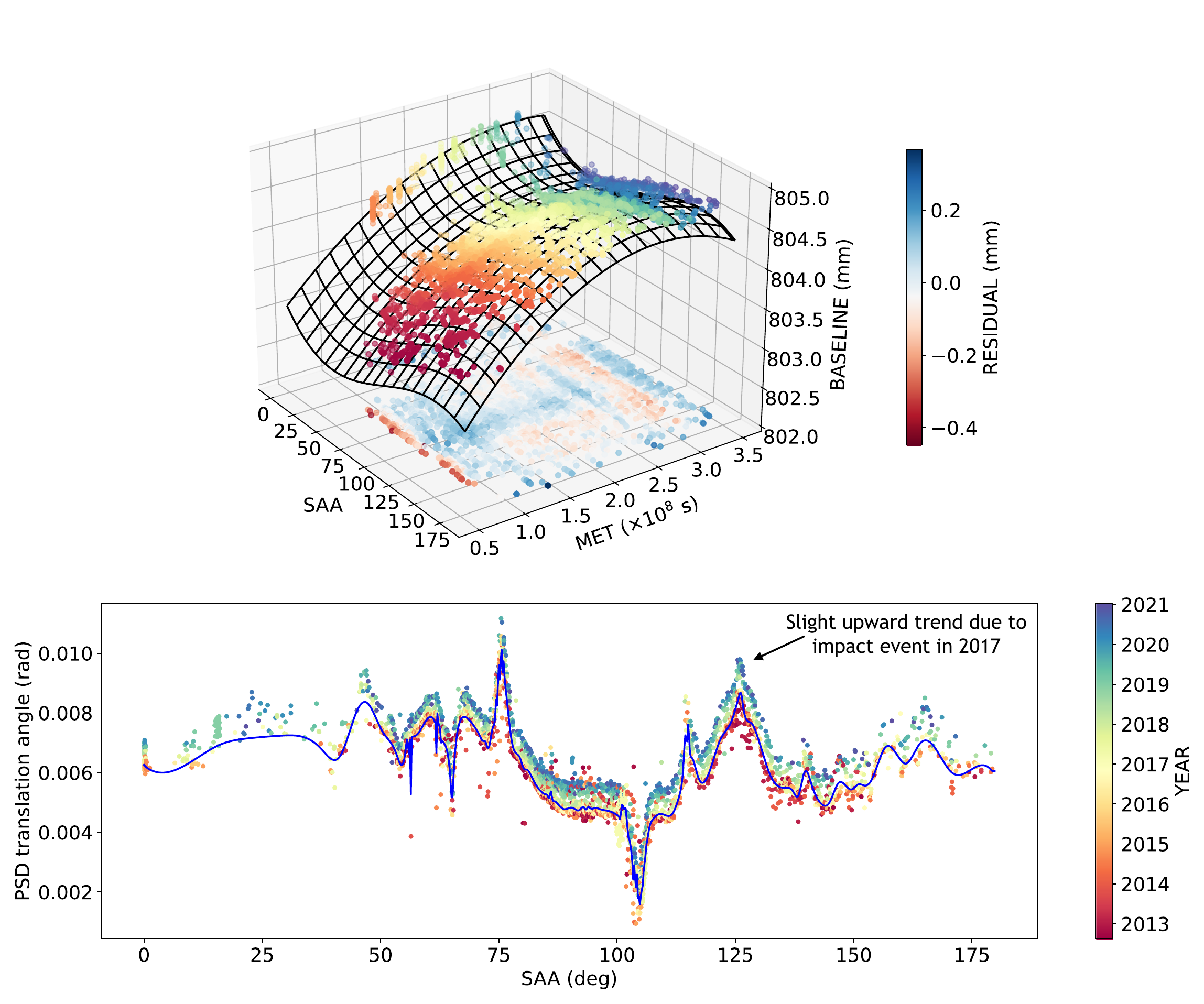}
\vspace{-4mm}
\end{center}
\caption 
{\label{fig:translationparams}
{\it Left,} the measured baseline between the median positions of the two PSD tracks in focal bench coordinates, plotted as a function of Mission Elapsed Time (MET) and SAA, and colored by time for clarity. The best-fitting 2D polynomial is shown in the form of a grid in black, and the residuals between the data and this model are shown beneath in blue and red. {\it Right,} the measured angle from the X axis between the two PSD tracks, plotted as a function of SAA and colored by time. The data prior to 2017 is fitted with a spline, plotted in blue.} 
\end{figure}

We find that the baseline and angle are both functions of SAA, with the baseline additionally a function of date (Fig.~\ref{fig:translationparams}). The baseline can be approximated with a third-order two-dimensional polynomial, which we fit to the data for the full duration of the mission to date. The variation in baseline is likely dominated by slight changes in the inclination of the lasers, where a variation of $\sim100$\,microradians corresponds to $\sim1$\,mm distance at the focal plane bench, and would be affected by long-term trends in the mast flexing, leading to the time dependence. The focal plane bench itself is made from aluminum and subject to thermal expansion amounting to $\sim$0.2\,mm on an orbital timescale, so while it may contribute to the scatter of the baseline relation, it is not a major factor in the baseline variation.

The angle is a complex function of SAA, with a slight upward trend in the angle after 2017 due to a known minor impact event in September of that year \cite{madsen20}. Given the effect of a slight change in angle is minimal (see below), we fit a spline curve to the data before 2017 and use that as our relation for predicting the angle.

While there is a small amount of scatter, the effect of changes in the value of baseline or angle at this scale is to move the derived sky positions of the photons, and therefore the entire resulting image, by a consistent small amount in sky coordinates (around 5 sky pixels per millimeter deviation). However, this does not have an effect on mapping the image to celestial coordinates, and therefore we are confident that these relations alone are sufficient for obtaining the parameters for the PSD track translation. 

In conclusion, it is possible to predict the position of the PSD1 track from the SAA and time of an observation. Using this information, the translated PSD0 track can replace the absent PSD1 track in the {\tt psdcorr} file, and used in the next steps of \nustar\ data reduction. 

\subsection{Mast Transform \& Quaternion Correction}
\label{sec:mast}

\begin{figure}
\begin{center}
\includegraphics[height=18cm]{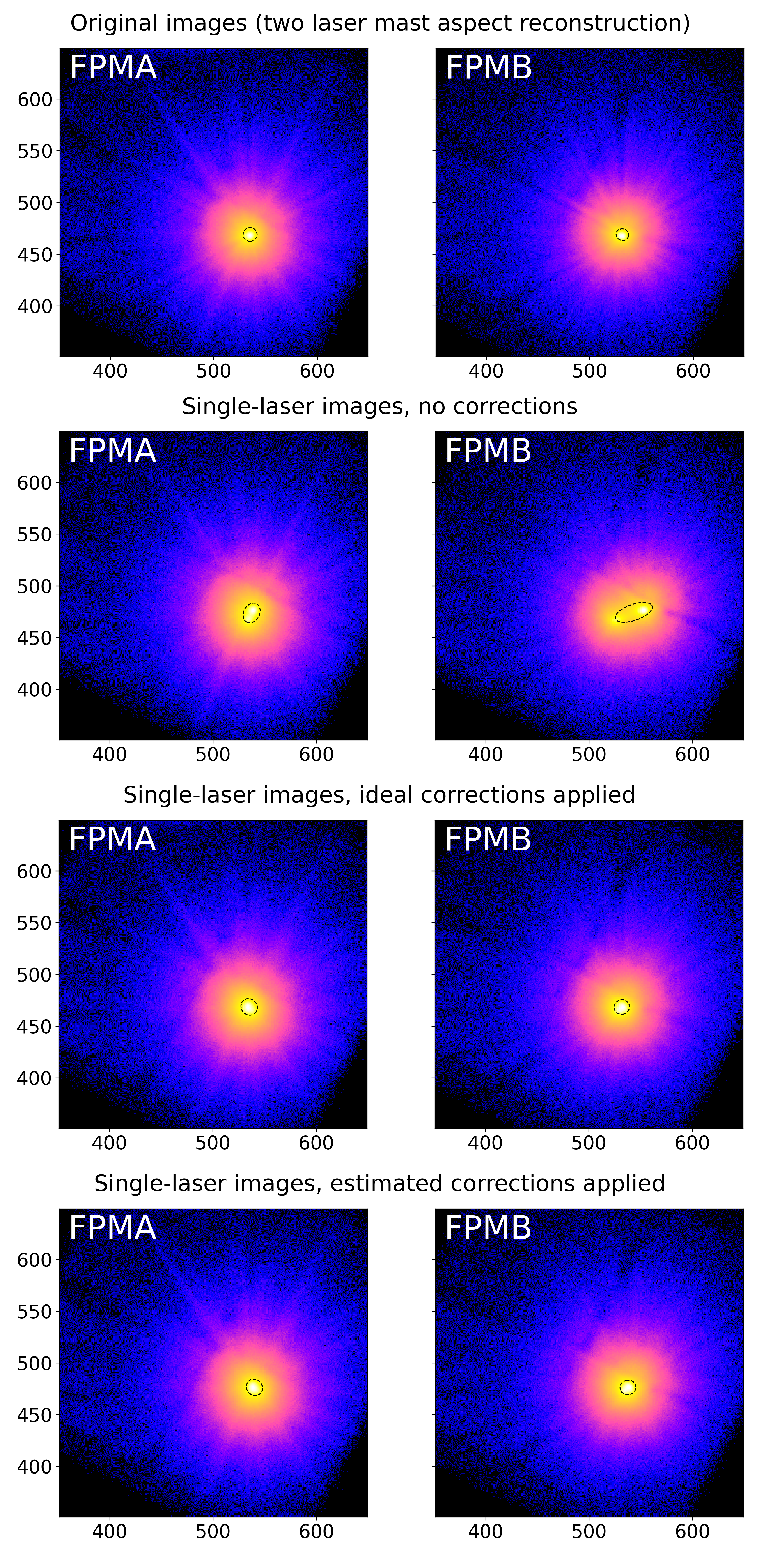}
\end{center}
\caption 
{\label{fig:psfs}
The \nustar\ PSF from observation 30301003002 after aspect reconstruction for FPMA {\it (left)} and FPMB {\it (right)}, in the following cases, from top to bottom: original aspect reconstruction with two functioning metrology lasers, aspect reconstruction with a single laser after only performing the PSD track translation step, aspect reconstruction with a single laser after performing the PSD track translation and applying ideal corrections to the {\tt mast} file, aspect reconstruction with a single laser after performing the PSD track translation and applying corrections to the {\tt mast} file estimated from spline fits to mast function parameters. The FWHMs of Gaussian fits to the PSFs are plotted with dashed black lines. The intensity is log-scaled to show the wings of the PSFs.} 
\end{figure}

The next step in the \nustar\ data reduction process is to run the {\tt numetrology} routine using the new {\tt psdcorr} file as an input, which produces the mast aspect solution and stores it in a new {\tt mast} file. This file is then used as an input to {\tt nupipeline} with {\tt runmetrology=no}, which turns off the automated metrology data reduction, in order to produce the event lists, which store the derived photon positions. We performed this for our example observation of GX~13+1, a bright source which allows us to clearly see the PSF shape. However, running {\tt nupipeline} on the new {\tt mast} file as-is yields a distorted PSF, particularly for FPMB (see Fig.~\ref{fig:psfs}). 

To discover the cause of this distortion, we compared the original {\tt mast} files with those produced by the single-laser approximation by plotting $T_x$, $T_y$, and $\theta$ against time in both cases (see Fig.~\ref{fig:masttracks}). We find that with a single laser track, most information regarding the two-dimensional transform is retained; the mean value and the amplitude of the variation is different, but the general shape of the curve over time is similar. However, all information about the mast twist angle motion is lost and is measured as a constant over the duration of the observation. In order to improve the single-laser reconstruction of the PSF, we need to adjust the single-laser {\tt mast} file so that the variations in the transform and twist angle more closely resemble those of the two-laser solution. 

To begin, we created an ideally corrected single-laser mast file by adjusting the mean and amplitude of the $T_x$ and $T_y$ curves over time to match those of the original {\tt mast} file, and fitting a simple sine function to the mast angle curve with the period fixed to the \nustar\ orbital period. To verify that this correction allows us to better reconstruct the PSF, we ran {\tt nupipeline} with the settings {\tt runmetrology=no} and {\tt inmastaspectfile} set to the corrected {\tt mast} file. This results in a PSF that is typically a little larger than the two-laser solution (by around 10\%), but no longer as distorted (see Fig.~\ref{fig:psfs}). Taking this level of fidelity as sufficient, we repeated this process for every observation in the \nustar\ archive with a bright point source ($>$50,000 counts) and recorded the following correction parameters: the amplitude of the $T_x$ and $T_y$ tracks, and the mean, amplitude, and phase of the best-fitting sine curve to the $\theta$ track. (The mean of the $T_x$ and $T_y$ tracks is functionally equivalent to the position of the translated PSD track -- that is, any difference from the original {\tt mast} file value is due to acceptable scatter in the estimation relation, so we do not consider these further.) We plotted these parameters against the SAA for each observation in order to see whether they can be predicted from the SAA.

\begin{figure}
\begin{center}
\includegraphics[width=16cm]{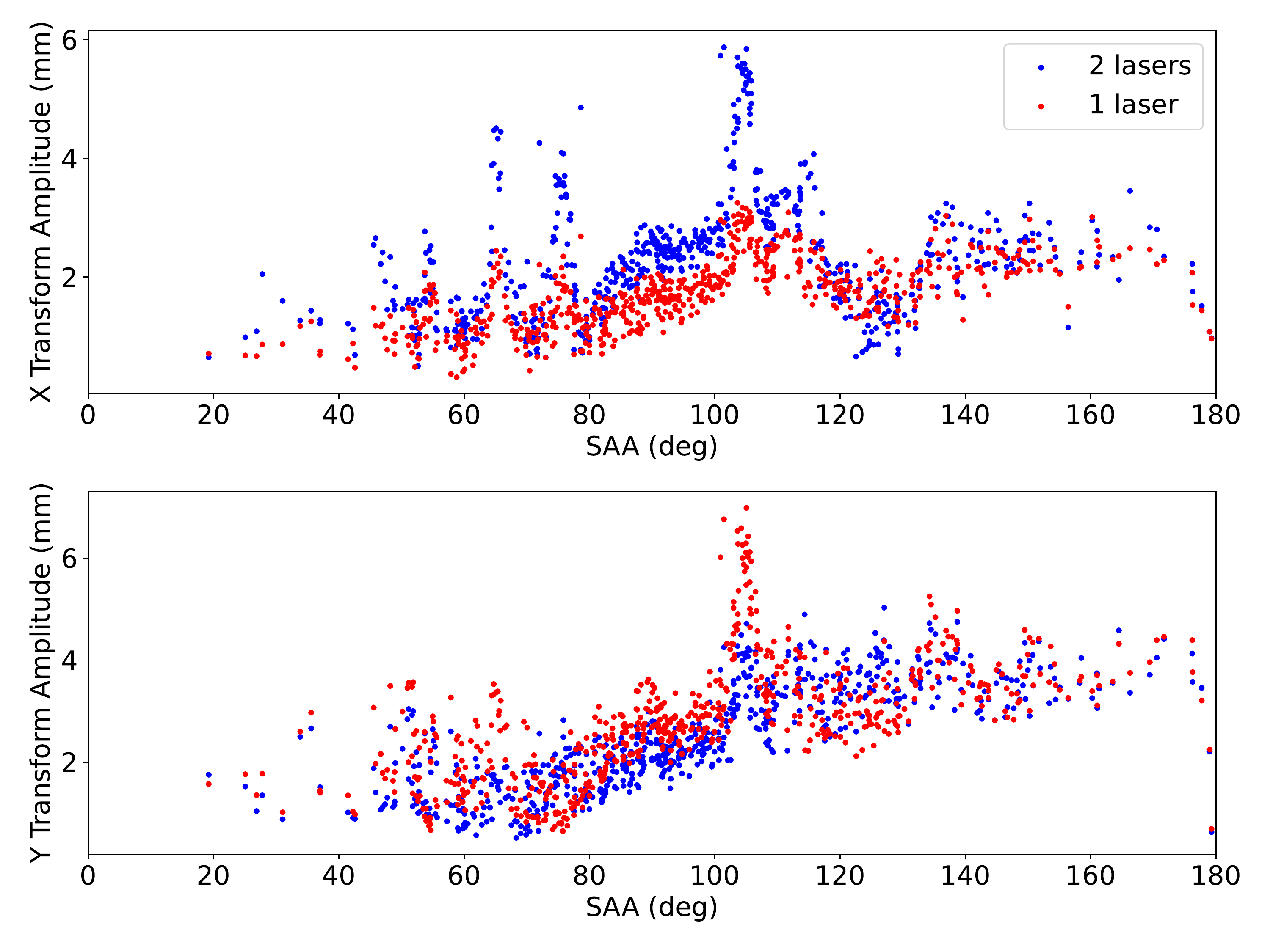}
\vspace{-4mm}
\end{center}
\caption 
{\label{fig:transamp}
The maximum amplitude of the variation in $T_x$ and $T_y$, for the original two-laser solution (blue) and the single-laser solution (red), plotted against SAA.} 
\end{figure}

\begin{figure}
\begin{center}
\includegraphics[width=16cm]{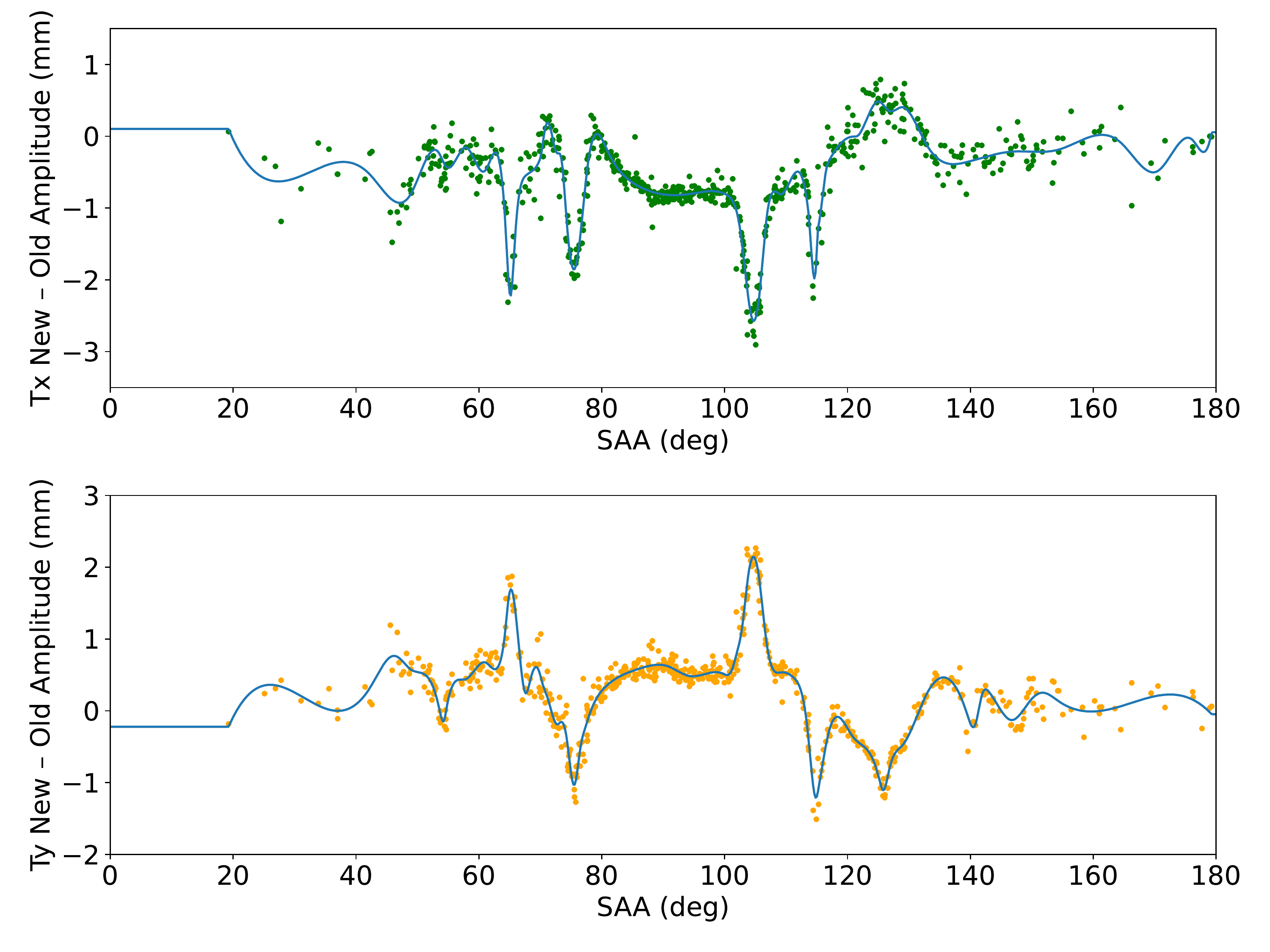}
\vspace{-4mm}
\end{center}
\caption 
{\label{fig:transampdiff}
The difference in maximum amplitude of the variation in $T_x$ (green) and $T_y$ (orange) between the single-laser solution and the original two-laser solution, plotted against SAA and fitted with a spline.} 
\end{figure}

First, we examined the amplitude of the variations in $T_x$ and $T_y$ as a function of SAA. Both the original amplitude and that of the single-laser solution vary approximately as a function of SAA, although there is a large amount of spread in the relation (Fig.~\ref{fig:transamp}). We found that a much tighter relation could be obtained if we plotted the difference between the new and old amplitudes (see Fig.~\ref{fig:transampdiff}). By fitting a spline to the data, we can use the SAA of an observation to estimate the amount to add to the single-laser $T_x$ and $T_y$ amplitudes to bring them to the value of the two-laser amplitudes.

\begin{figure}
\begin{center}
\includegraphics[width=16cm]{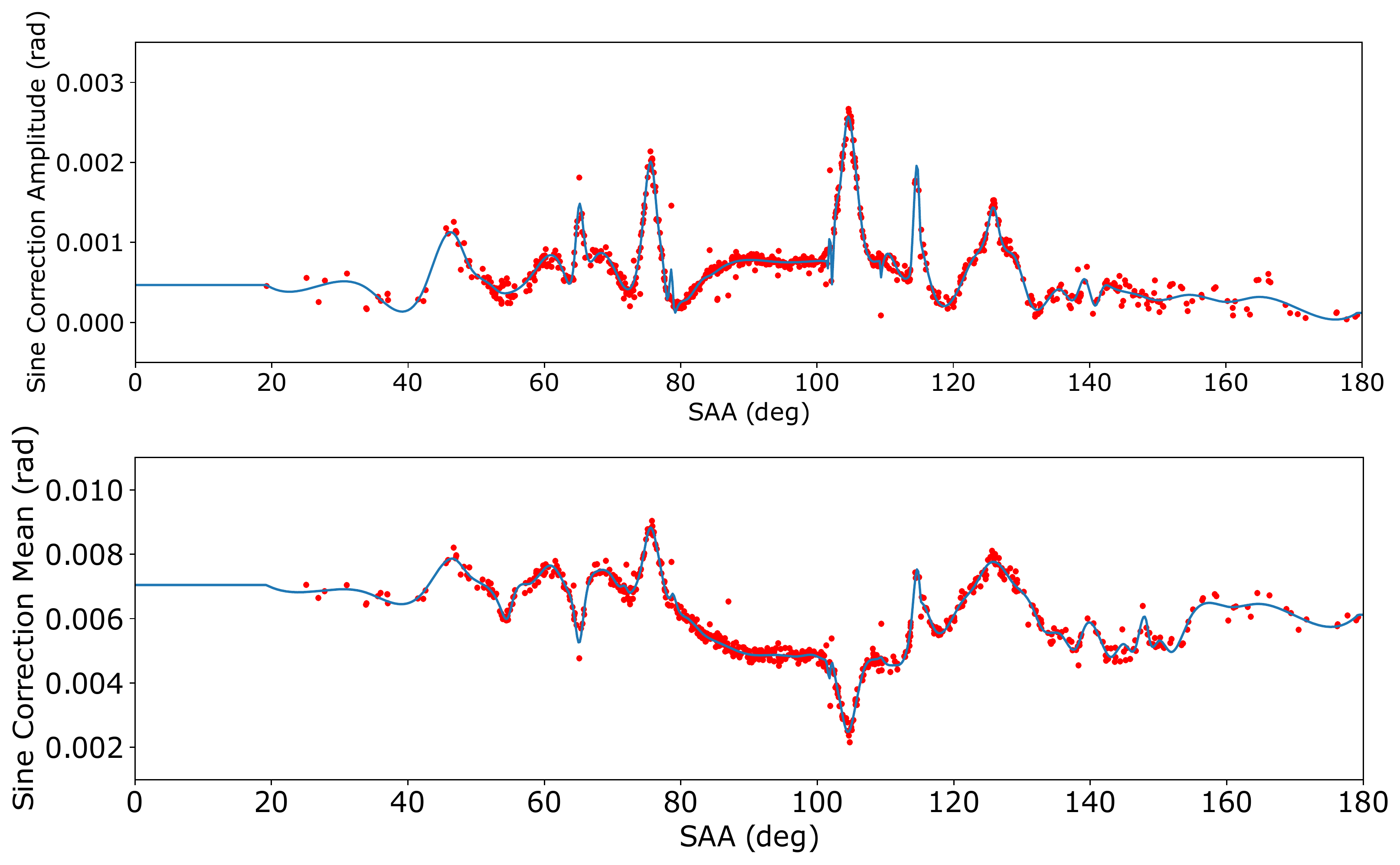}
\end{center}
\caption 
{\label{fig:sineparams}
{\it Top,} the amplitude of the best-fitting sine curve to the mast twist angle motion over time, plotted against SAA and fitted with a spline. {\it Bottom,} the mean of the best-fitting sine curve to the mast twist angle motion over time, plotted against SAA and fitted with a spline.} 
\end{figure}

\begin{figure}
\begin{center}
\includegraphics[width=16cm]{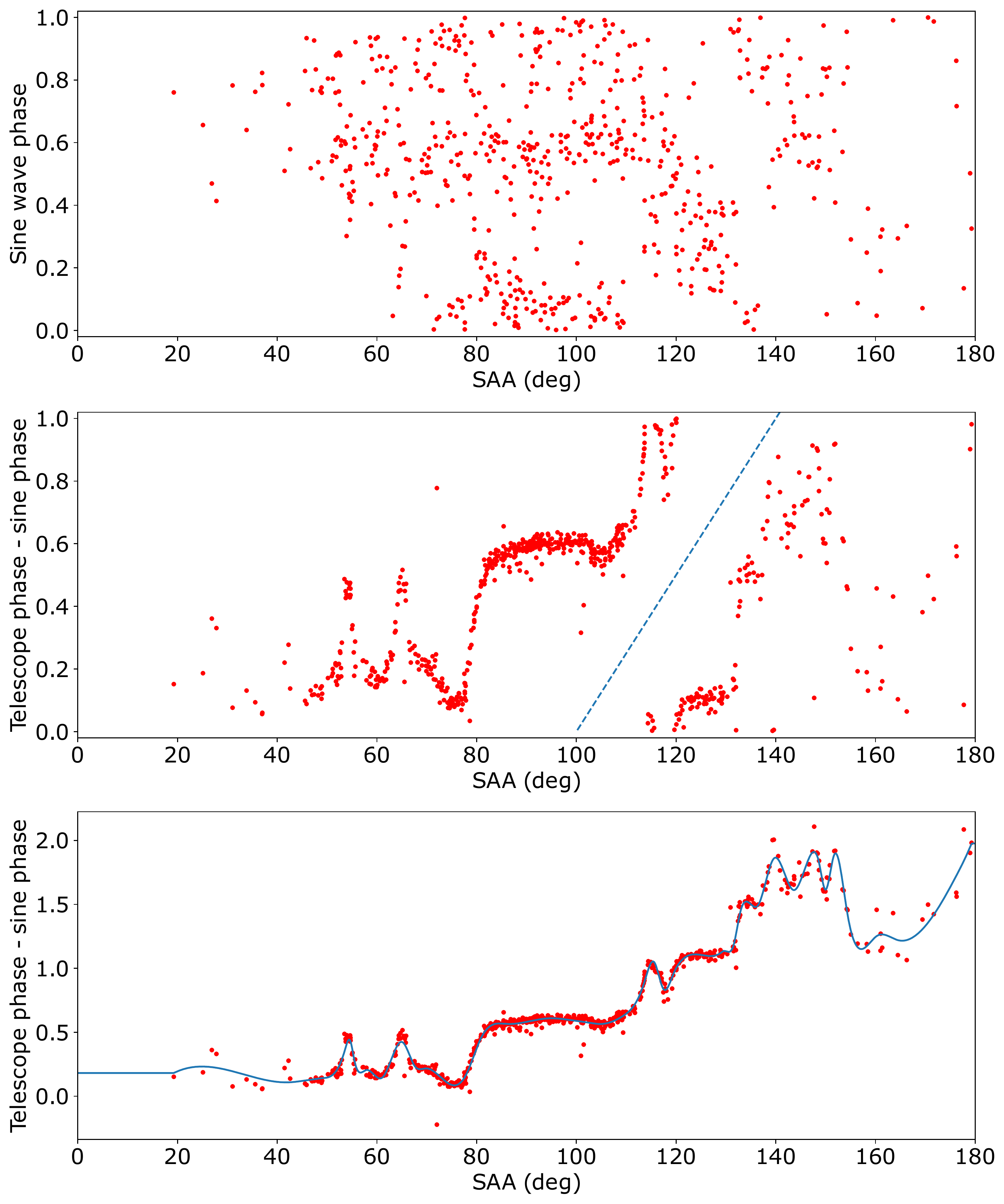}
\end{center}
\vspace{-2mm}
\caption 
{\label{fig:phase}
{\it Top,} the phase of the best-fitting sine curve to the mast twist angle motion over time, plotted against SAA. {\it Middle,} the difference between the telescope phase as measured from first emergence into daylight, and the phase of the mast angle modulation, plotted against SAA. Points below the dashed line can be incremented by 1 to create a continuous curve in phase. {\it Bottom,} as Middle, with the indicated points incremented by 1 (and 2 in a few cases), and the resulting curve fitted with a spline.} 
\end{figure}

The mean and amplitude of the sine curve that best fits the mast twist angle variation are both a relatively tight function of SAA, and can simply be fitted with a spline to be able to estimate these parameters directly from the SAA of an observation (Fig.~\ref{fig:sineparams}). The phase of the sine curve does not vary with a simple function of SAA (Fig.~\ref{fig:phase}, top), so we compared this phase with that of the orbit phase of the telescope as measured from the time it emerges into sunlight from behind the Earth's shadow (whether or not the telescope is in sunlight is recorded in the DAY column of the Orbit file). We found that the difference between these two phases is a function of SAA, albeit not continuous (Fig.~\ref{fig:phase}, middle). Since phase is a periodic quantity, we added to selected data points intervals of 1 to bring them into a single continuous function, which we fitted with a spline curve (Fig.~\ref{fig:phase}, bottom). At this point, we have successfully demonstrated that all information for a sine approximation to the mast twist angle motion can be obtained from the observation SAA and orbit data. 

Therefore, with no knowledge of the contents of the two-laser {\tt mast} file, it can be approximated by taking the single-laser {\tt mast} file and correcting it by adjusting the transform variation amplitudes and replacing the constant mast twist angle with a sine wave approximation, the parameters for which can all be derived from the observation SAA and the Orbit file.

\section{Results}
\label{sec:results}

In order to determine whether this method can successfully reconstruct the PSF for all observations, and how well it does so, we performed both steps of the process -- estimating the PSD translation parameters from the SAA and observation date, and the {\tt mast} file adjustment parameters (i.e. the differences to apply to the $T_x$ and $T_y$ transforms and the parameters for the sine curve approximation to the $\theta$ variation) from the SAA and orbit -- on all observations of a single bright point source ($>$50,000 counts). We generated new {\tt psdcorr} and {\tt mast} files, corrected the {\tt mast} files using the derived adjustment parameters, then ran {\tt nupipeline} using the corrected {\tt mast} file as an input for the metrology. We created intensity images from the resulting event files and fitted them with a 2D Gaussian model which, while the \nustar\ PSF is a more complex and peaked shape than a Gaussian, still provides a good indication of whether the PSF is extended or elongated compared with the PSF of the two-laser scenario. 

\subsection{General PSF fitting results}
\label{sec:psf}

\begin{figure}
\begin{center}
\includegraphics[width=16cm]{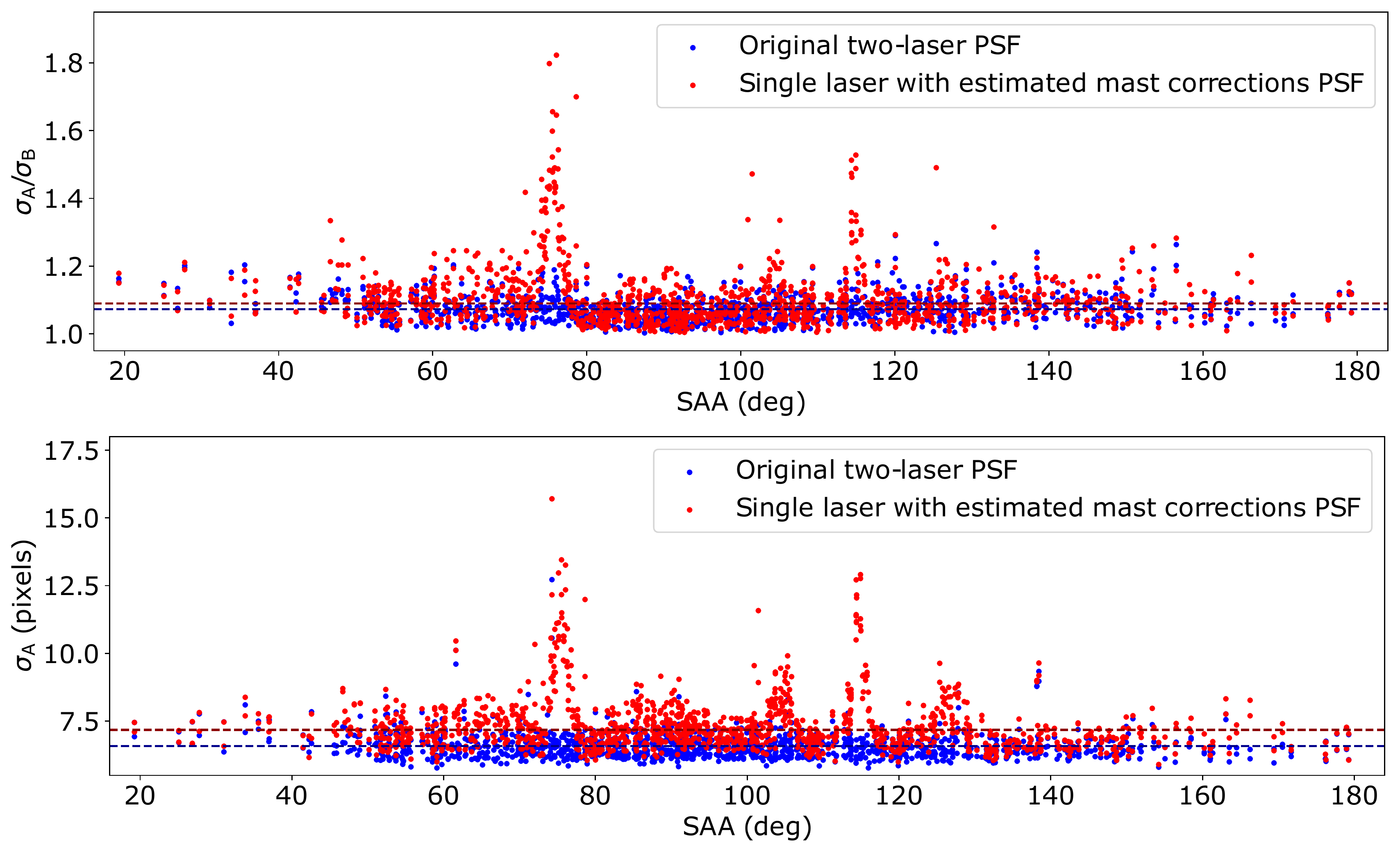}
\vspace{-6mm}
\end{center}
\caption 
{\label{fig:results}
{\it Top,} the ratio of the semimajor to semiminor axis of the 2D Gaussian fit to the PSF. {\it Bottom,} the size of the semimajor axis of the 2D Gaussian fit to the PSF in pixels, where each pixel covers an angular size of 2.46\,arcsec. Fits to the original PSF reconstructed using the two-laser mast solution are plotted in blue, and fits to the PSF reconstructed using the single-laser approximation are plotted in red, with mean values (calculated ignoring data with SAA = 74--78 and 114--116) are plotted in dark blue and dark red respectively. Only observations for which a good fit was obtained are plotted (see text).} 
\end{figure}

We plot the ratio of the FWHM semimajor to semiminor axis of the 2D Gaussian fit to the PSF against the SAA for each observation, both for the single-laser reconstructed images and the original images generated using the two-laser solution, to see how well this single-laser approximation method reproduces the two-laser results across different SAAs (Fig.~\ref{fig:results}, top). We only plot observations for which a successful automated Gaussian fit could be made to the data, which we define to have a semimajor axis less than 20 pixels (49.2\,arcsec). Bad fits are often due to multiple bright sources in the field of view or the presence of stray light. We also filter out fits for which the original PSF is elongated with a major-to-minor axis ratio greater than 1.3, usually due to the source being located at a high off-axis angle. 

We find the elongation of the PSF for the single-laser approximation to be very similar to that of the two-laser solution except for two major spikes at SAA $\approx76$ and SAA $\approx115$. If we ignore observations with SAA in the ranges 74--78 and 114--116, we find the mean axis ratio for the original two-laser solution to be $1.07\pm0.04$, and the mean ratio for the single-laser approximation to be $1.11\pm0.09$. In terms of the elliptical distortion of the PSF, the single-laser approximation performs almost as well as the two-laser solution except for two small ranges of SAA.

If we plot the size in pixels of the semimajor axis of the Gaussian fit by SAA (Fig.~\ref{fig:results}, bottom), there are two major peaks which correspond to those where the PSF is elongated,  at SAA $\approx76$ and SAA $\approx115$. There are also additional, smaller peaks at SAA $\approx105$ and SAA $\approx125$, and a general excess at SAAs between 80 and 95. Ignoring the same peaks as before, we find that the mean semimajor axis of fit to the two-laser PSF is $6.6\pm0.5$ pixels, or $16.2\pm1.2$\,arcsec (this is close to the actual PSF FWHM of 18\,arcsec, which is smaller than the HPD due to the peaked shape of the PSF). The mean semimajor axis for the single-laser PSF is $7.4\pm1.0$ pixels, or $18.1\pm2.6$\,arcsec. 

\begin{figure}
\begin{center}
\includegraphics[width=16cm]{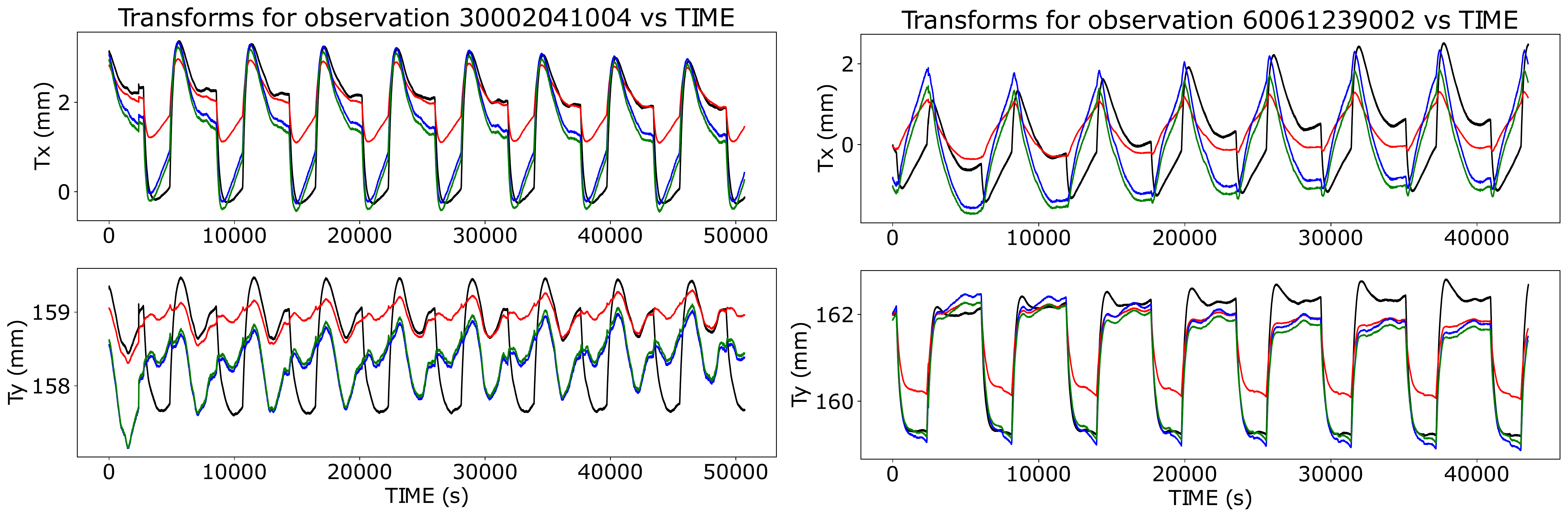}
\end{center}
\caption 
{\label{fig:badsaa}
The X and Y transforms for observations 30002041004, of SS~433 at SAA = 75.16 {\it (left)}, and 60061239002, of NGC~4992 at SAA = 114.42 {\it (right)}. Colors as in Fig.~\ref{fig:masttracks}. The estimated track (green) is not a good approximation to the original track (black) for $T_y$ in the case of 30002041004, and for $T_x$ in the case of 60061239002.} 
\end{figure}

Therefore the reconstructed PSF using a single laser is systematically a little larger than that of the two-laser solution, but only severely distorted at two narrow ranges of SAA around SAA = 76 and SAA = 115. 

We investigate observations in these SAA ranges to determine what is causing the high level of distortion. We find that at these particular SAAs, the periodic variation in the Y transform for SAA $\approx$ 76 and X transform for SAA $\approx$ 115 cannot be replicated well by the mast solution derived from the transposed PSD track, even when its amplitude is adjusted, due to a double-peaked or otherwise complex shape to the variation that is not retained in a single PSD track (Fig.~\ref{fig:badsaa}).

\begin{figure}
\begin{center}
\includegraphics[width=16cm]{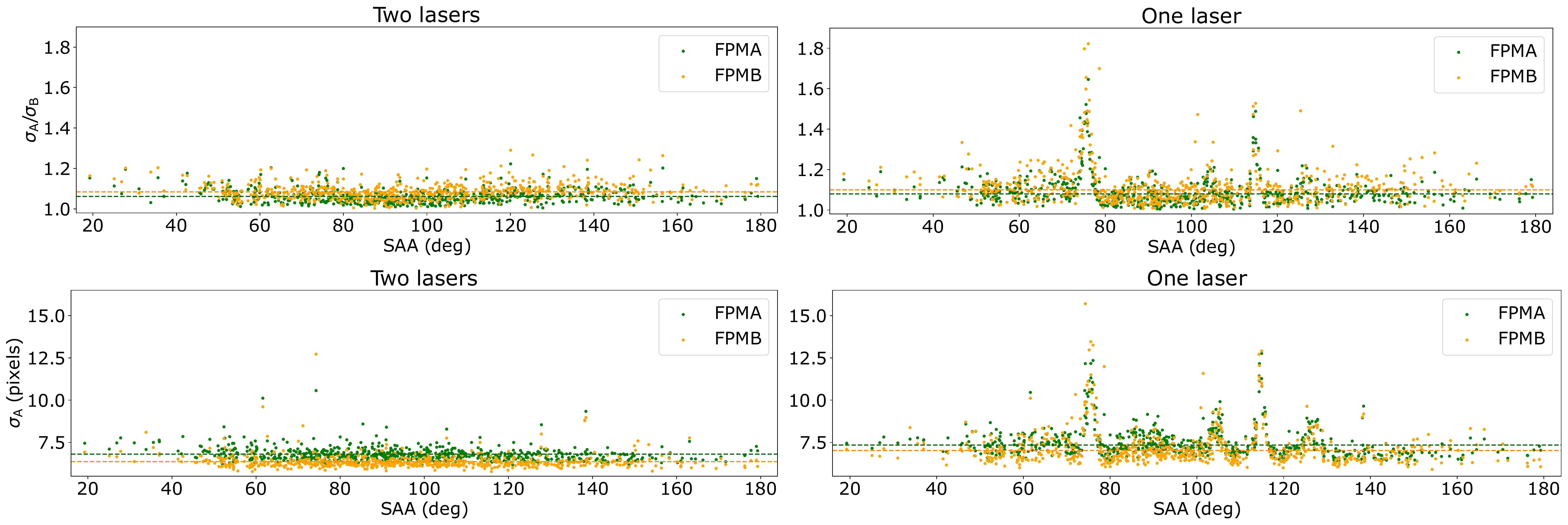}
\vspace{-4mm}
\end{center}
\caption 
{\label{fig:resultsfpm}
{\it Top,} the ratio of the semimajor to semiminor axis of the 2D Gaussian fit to the PSF. {\it Bottom,} the size of the semimajor axis of the 2D Gaussian fit to the PSF in pixels, where each pixel covers an angular size of 2.46\,arcsec. {\it Left,} the results for the original two-laser mast aspect reconstruction. {\it Right,} the results for the single-laser mast aspect reconstruction. The FPMA results are plotted in green, and the FPMB results are plotted in orange. Mean values (calculated ignoring data with SAA = 74--78 and 114--116) are plotted in dark green and dark orange respectively.} 
\end{figure}

We also examine whether there is any significant difference between the results for the FPMA and FPMB detectors (Fig.~\ref{fig:resultsfpm}). We find that while there is a slight systematic difference between FPMA and FPMB, this is fairly consistent with the difference seen in the original two-laser solution, so there is no additional discrepancy between the two detectors introduced by the single-laser reconstruction method.

\subsection{Her X-1}
\label{sec:herx1}

In order to more closely examine changes to the PSF between the two-laser and one-laser mast aspect reconstruction methods, we performed a detailed study of five observations of Her~X-1 taken at different dates and SAAs, comparing the data products generated using the two-laser solution to those generated using the single-laser approximation. We present the observations used in Table~\ref{tab:herx1}. Her X-1 has been used as a PSF calibrator for many X-ray observatories due to its brightness and low $n_H$, which ensures that there is no dust scattering halo extending the PSF.

\begin{table}[ht]
\caption{The \nustar\ observations of Her~X-1 studied for this investigation.} 
\label{tab:herx1}
\begin{center}       
\begin{tabular}{|c|c|r|} 
\hline
\rule[-1ex]{0pt}{3.5ex}  Observation ID & Date & SAA  \\ %Also max diff in EEF?
\hline\hline
\rule[-1ex]{0pt}{3.5ex} 30002006002 & 2012-09-19 & 79  \\
\rule[-1ex]{0pt}{3.5ex} 30002006005 & 2012-09-22 & 77  \\
\rule[-1ex]{0pt}{3.5ex} 30402009004 & 2019-06-23 & 119  \\
\rule[-1ex]{0pt}{3.5ex} 30602003002 & 2020-08-12 & 98  \\
\rule[-1ex]{0pt}{3.5ex} 90102002002 & 2015-08-03 & 103  \\
\hline 
\end{tabular}
\end{center}
\end{table} 

\begin{figure}
\begin{center}
\includegraphics[width=16cm]{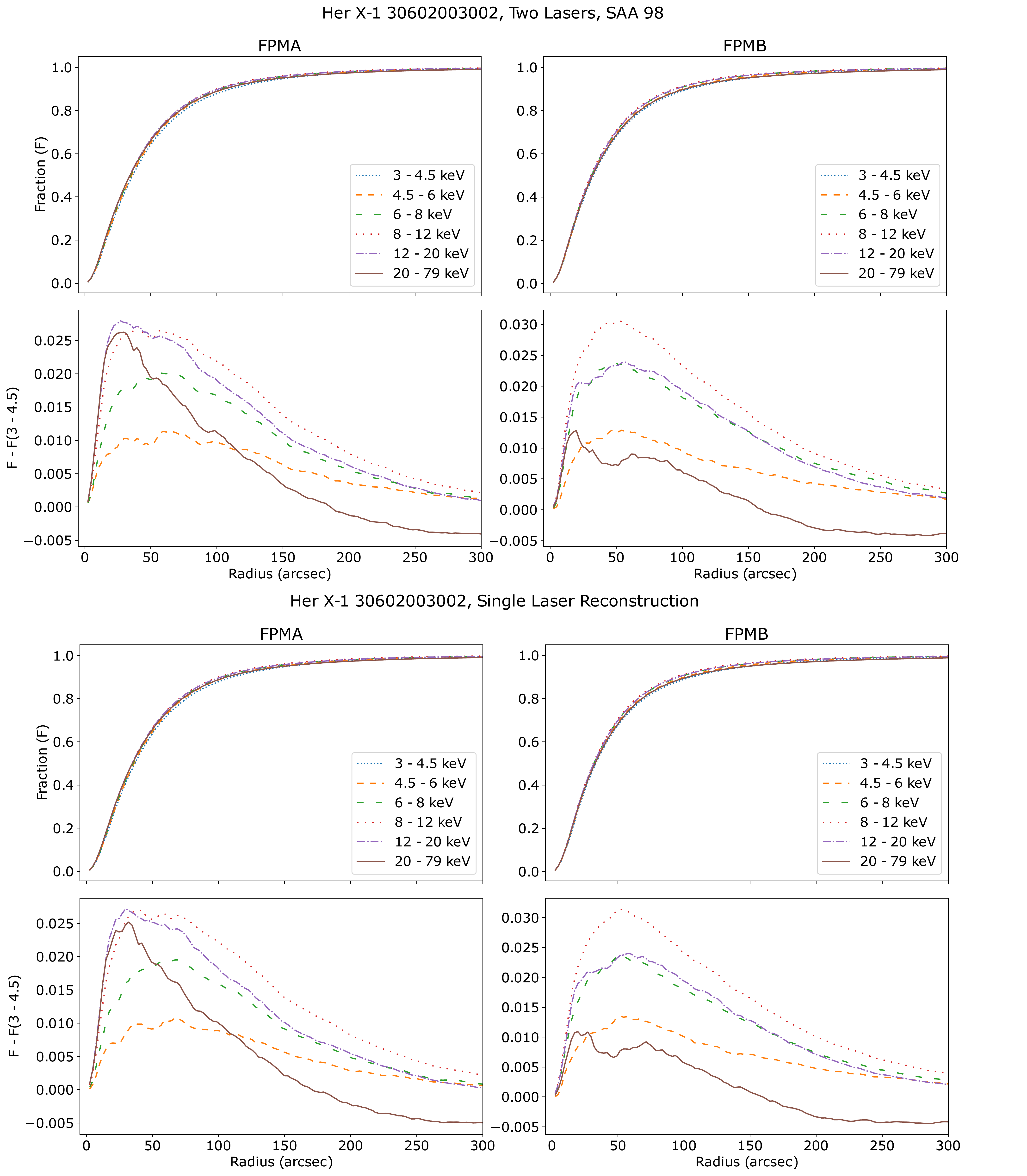}
\end{center}
\caption 
{\label{fig:herx1eef}
Enclosed energy functions for Her~X-1, from observation 30602003002. {\it Top panel,} EEFs for the original two-laser solution. {\it Bottom panel,} EEFs for the single-laser reconstruction. In each panel, the results for FPMA are plotted on the left, and FPMB on the right. The upper plots show the EEFs in each energy band, as a fraction of the total enclosed energy, and the lower plots shows the differences between the EEFs in each band and the EEF in the 3--4.5\,keV energy band.} 
\end{figure}

For each observation, we ran through the entire process detailed in Section~\ref{sec:method} to produce single-laser reconstructions of the images and studied the shape of the PSF compared to that of the two-laser solution. Following a method based on An et al. (2014) \cite{an14}, we modeled the background using {\tt nuskybgd} \cite{wik14} and subtracted it from the image of Her~X-1. Then we found the centroid of the PSF and measured the enclosed energy function (EEF) of the PSF within circular apertures with radii at 1-pixel intervals up to a radius of $\sim$10 arcminutes (240 pixels), over the following energy bands: 3--4.5\,keV, 4.5--6\,keV, 6--8\,keV, 8--12\,keV, 12--20\,keV, and 20--79\,keV. We also plotted the difference between the EEF in each band and the lowest-energy band (see Fig.~\ref{fig:herx1eef}).

\begin{figure}
\begin{center}
\includegraphics[width=16cm]{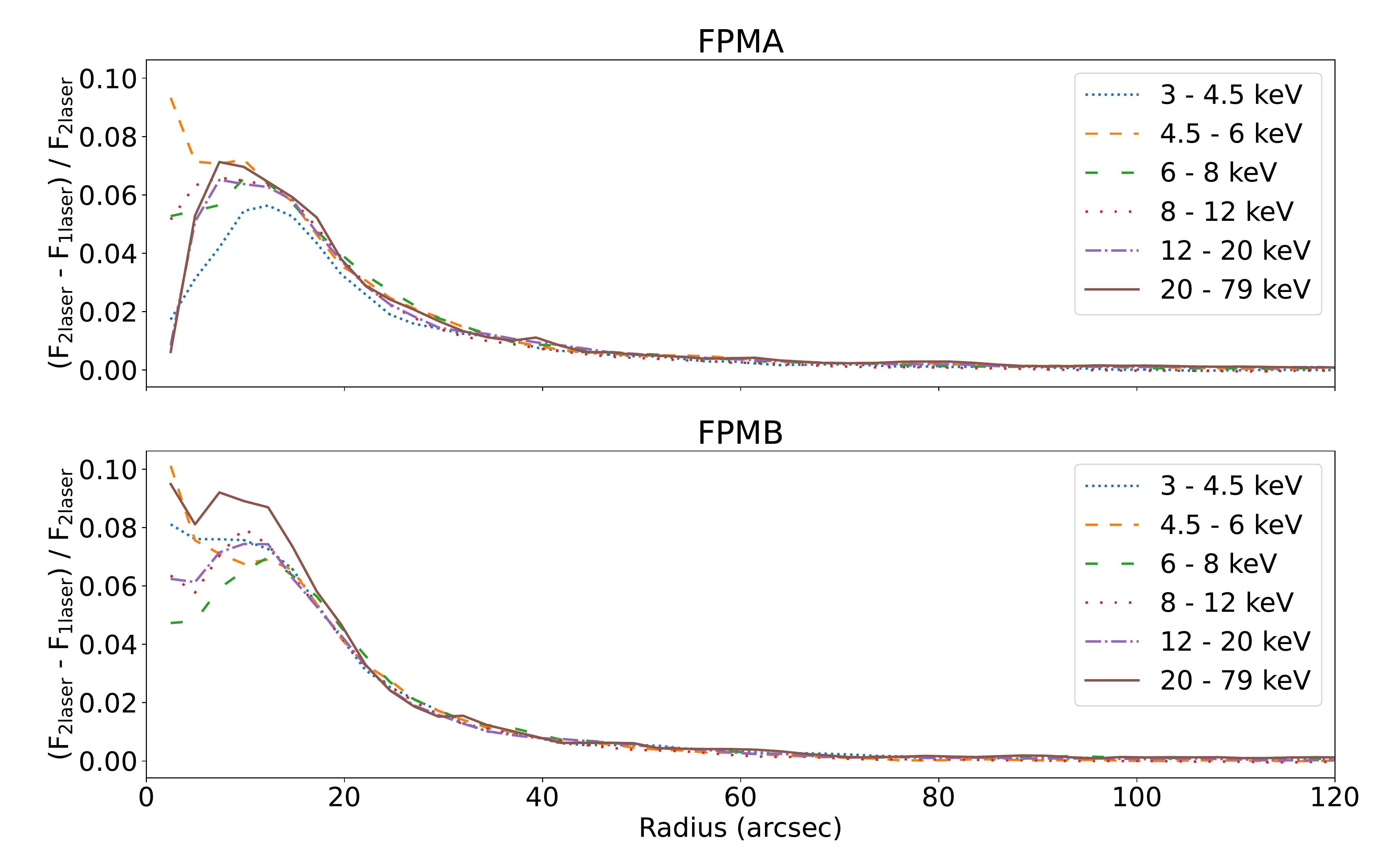}
\end{center}
\caption 
{\label{fig:herx1eefdiff}
The fractional difference between the two- and one-laser scenarios as a function of radius, for observation 30602003002. Line colors and styles are the same as in Fig.~\ref{fig:herx1eef}.} 
\end{figure}

We see that the PSF tends to be more peaked at higher energies, up to the point that our imperfect background modeling becomes significant, and there is very little difference in the PSF shape between the one- and two-laser cases. We can investigate the difference between the two scenarios more closely by plotting the fractional difference between the two-laser and one-laser EEFs at each radius (Fig.~\ref{fig:herx1eefdiff}). For observation 30602003002 we can see that, within the first 50\,arcseconds, the enclosed energy fraction for the two-laser solution is higher than that for the single-laser approximation, meaning that the PSF reconstructed with just one laser is a little more blurred out than the original PSF, consistent with our results from Section~\ref{sec:psf} showing a slightly larger PSF size for reconstruction using a single laser. The maximum fractional difference is $\sim$10\% at a radius of about 10\,arcseconds, and with a region of radius greater than $\sim$50\,arcseconds, the difference in enclosed energy is minimal, at $<$1\%. The results for the other observations are qualitatively similar, with 5--10\% maximum difference between the one- and two-laser scenarios being typical, and the difference becoming minimal by 50\,arcseconds from the centroid.

\begin{figure}
\begin{center}
\vspace{-6mm}
\includegraphics[width=16cm]{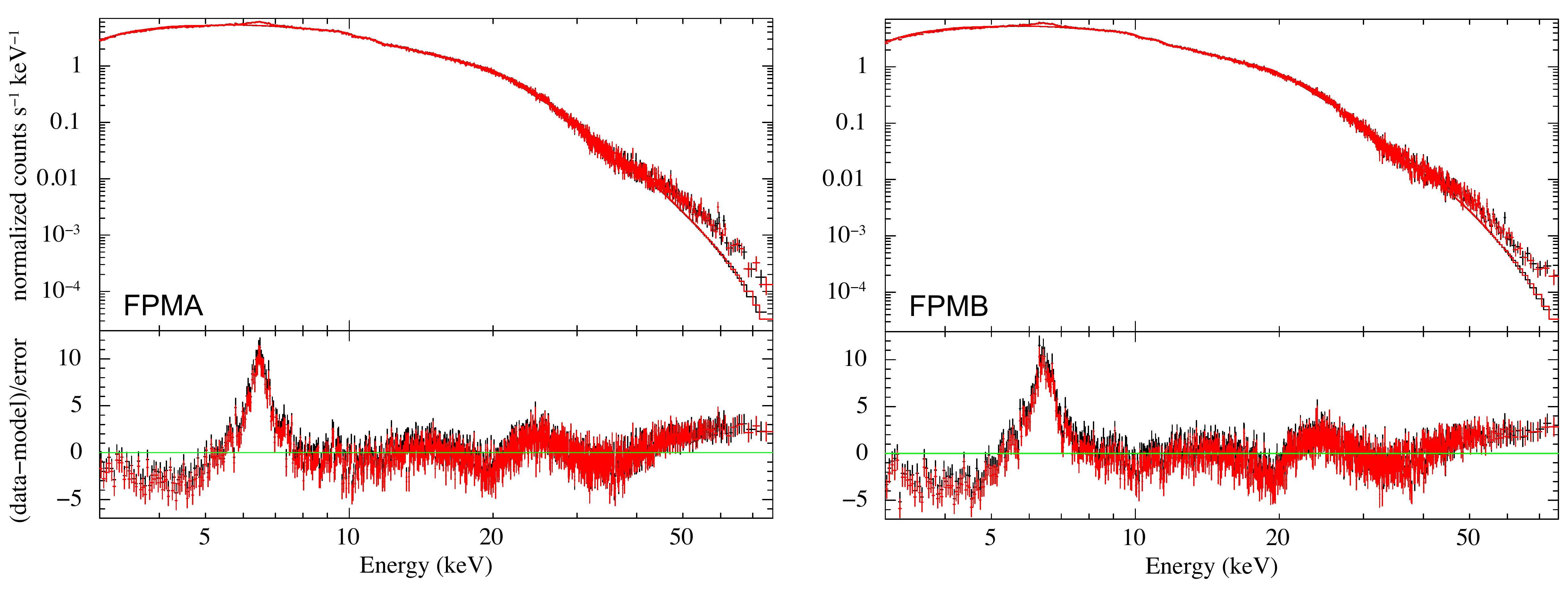}
\end{center}
\caption 
{\label{fig:herx1spec}
The spectrum of Her~X-1 extracted from a 50\,arcsecond radius and fitted with a power-law with a high-energy cut-off, along with the residuals from the model, for FPMA (left) and FPMB (right). The spectrum from the original two-laser PSF is plotted in black, and the spectrum from the single-laser reconstruction is plotted in red.} 
\end{figure}

We can examine any differences in energy another way by looking at the spectrum extracted from the original PSF and that reconstructed from a single laser, using {\tt nuproducts} to extract the spectrum and associated response files. We plot the spectrum of Her~X-1 extracted from a source region of 50\,arcseconds (subtracting background extracted from a 100\,arcsecond region as far as possible from the source location), along with the residuals from fitting it with a simple power-law model with a high-energy cutoff (Fig.~\ref{fig:herx1spec}). We can see that the difference between the one- and two-laser scenarios is negligible compared even with the difference between the FPMA and FPMB detectors, and spectral fit parameters are the same to within $<1$\%. This also applies to the rest of the observations, at extraction radii between 50\,arcseconds and 4\,arcminutes, and with more sophisticated models to fit the various features of the spectrum.

\begin{figure}
\begin{center}
\includegraphics[width=16cm]{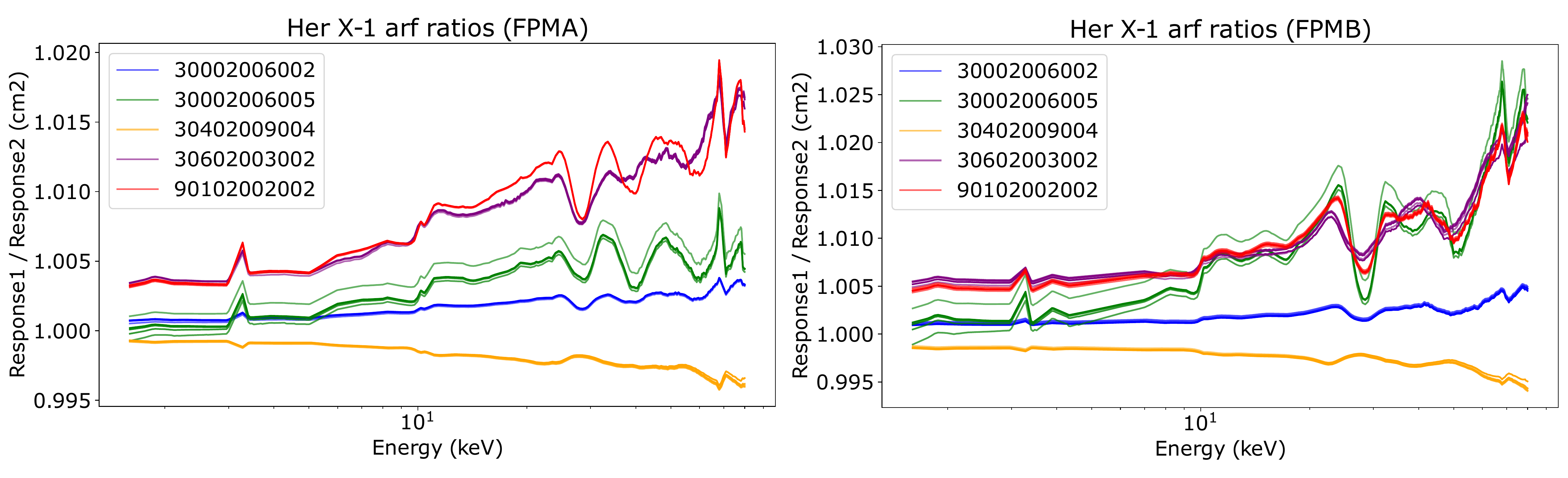}
\end{center}
\caption 
{\label{fig:herx1arf}
The ratio between the ARFs generated from the single-laser products and the two-laser products, for FPMA (\textit{left}) and FPMB (\textit{right}). Ratios for different source extraction radii are plotted as different shades of the color for each observation, though most of these lines are overlapping.} 
\end{figure}

Finally, we also examine the ancillary response functions (ARFs) generated for this source. For each observation, we plot the ratio between the two- and one-laser ARFs as a function of energy, at a range of extraction radii between 50\,arcseconds and 4\,arcminutes (Fig.~\ref{fig:herx1arf}). We find that different extraction radii do not, for the most part, affect the ratio between the ARFs. The ratio is low, within $\lesssim0.5$\%, at low energies, and increases with energy, up to a maximum of 2--3\% at high energies. This is within the calibration errors at the respective energies \cite{madsen15}.

In conclusion, we do not expect there to be any significant difference in the data products extracted from source regions greater than 50\,arcseconds for images reconstructed from our single-laser approximation method.

\section{Changing the laser used for reconstruction}
\label{sec:reverse}

\begin{figure}
\begin{center}
\includegraphics[width=16cm]{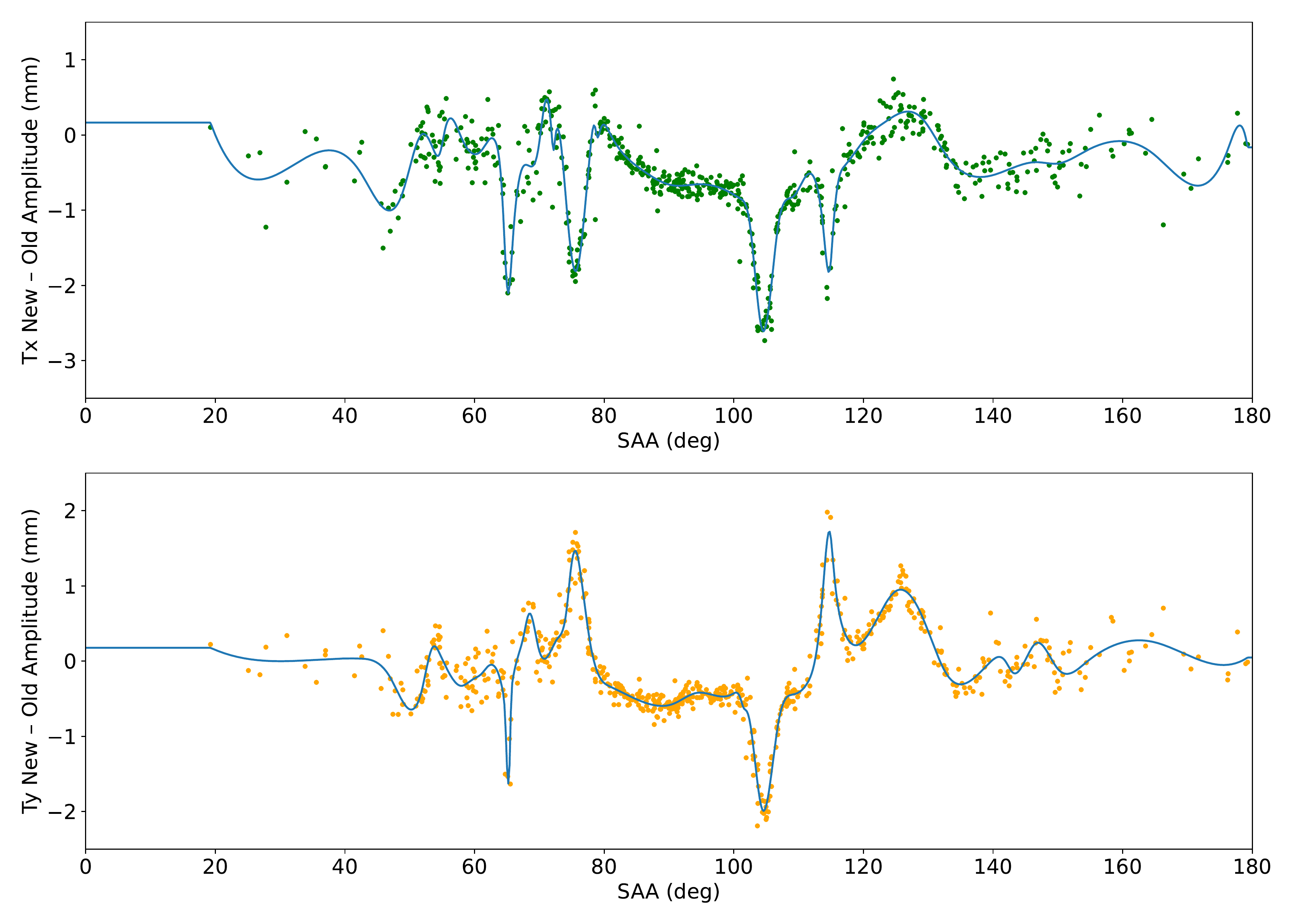}
\end{center}
\caption 
{\label{fig:transampdiff_reverse}
The difference in maximum amplitude of the variation in $T_x$ and $T_y$ between the single-laser solution and the original two-laser solution, using LASER1 as a basis rather than LASER0, plotted against SAA and fitted with a spline. Colors are as in Fig.~\ref{fig:transampdiff}.} 
\end{figure}

The previous analysis assumes the scenario that LASER1 is the one to fail and LASER0 is operational, and thus uses the LASER0 track to reconstruct the mast aspect solution. We used this assumption because, of the two lasers, the intensity of LASER1 has been declining faster than that of LASER0, and we expect that, should a laser fail in the future, it is more likely to be LASER1. However, we still investigated the opposite scenario for completeness, assuming that LASER0 fails and performing the previously-described process in a similar manner using LASER1: we used the same estimated baseline and angle (adding $\pi$ to reverse the translation direction) to make a copy of the PSD1 track at the position of the PSD0 track, and produced a new {\tt mast} file. Since the parameters of the sine wave approximation to the mast twist angle do not depend on the PSD tracks at all, these estimators remain the same. The only two parts that we need to regenerate for this case are the differences in the amplitudes of $T_x$ and $T_y$, which we show in Fig.~\ref{fig:transampdiff_reverse}. The SAA relation for the $T_x$ amplitude difference is very similar regardless of which laser is used, and the relation for $T_y$ is close to the inverse of that for using LASER0 for mast aspect reconstruction.

\begin{figure}
\begin{center}
\includegraphics[width=16cm]{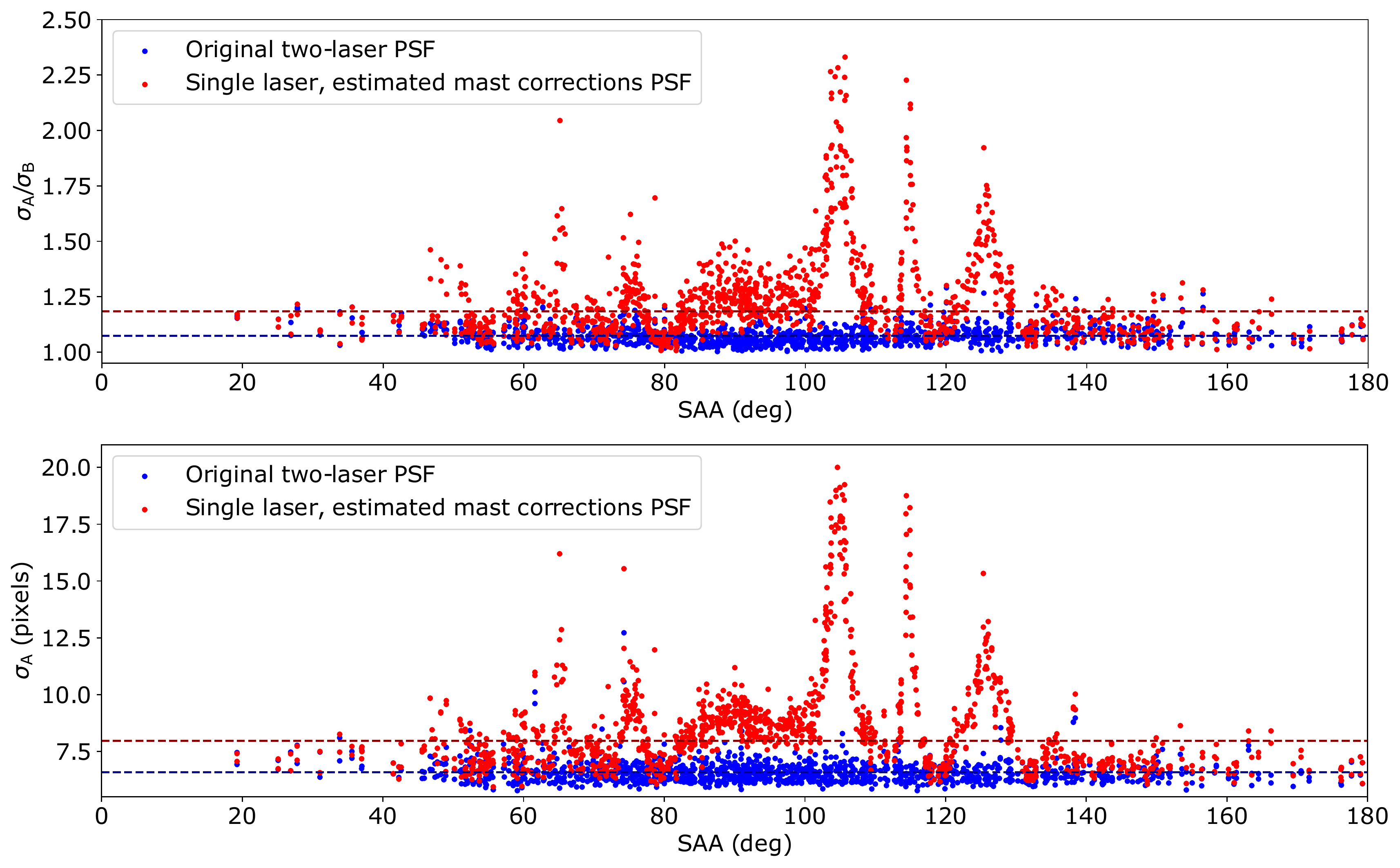}
\end{center}
\caption 
{\label{fig:resultsreverse}
Results for the scenario of using LASER1 to reconstruction the mast aspect solution.  {\it Top,} the ratio of the semimajor to semiminor axis of the 2D Gaussian fit to the PSF. {\it Bottom,} the size of the semimajor axis of the 2D Gaussian fit to the PSF in pixels. Lines and colors as in Fig~\ref{fig:results}, with mean values calculated ignoring data with SAA = 101--107, 114--116, and 124--128.} 
\end{figure}

When we make these adjustments to the {\tt mast} file, run the pipeline, and fit the resulting images with a 2D Gaussian as in Section~\ref{sec:psf}, we find that the average elongation of the PSF (not counting the three largest peaks) is $1.24\pm0.2$, greater than for the scenario in which we use LASER0, with three large peaks in PSF elongation at SAA $\approx$ 104, $\approx$115, and $\approx$126, and several other ranges of SAA with greater PSF distortion than the two-laser scenario (Fig.~\ref{fig:resultsreverse}). In this case, increases in the semimajor axis always drive increases in the distortion, with the semiminor axis not increasing to keep the PSF roughly circular as it does in some cases for the scenario in which we use LASER0. In summary, the failure of LASER0 would lead to a generally worse scenario than the failure of LASER1. 

This asymmetry between how well we can reconstruct the PSF using the PSD0 track versus the PSD1 track is unsurprising, as we would not expect \nustar's mast motion to be exactly symmetrical in terms of the extent to which it affects each side of the spacecraft. This may also indicate that the location of either the laser or the optic has a small error leading to the larger reconstruction error. Fortunately, it happens to be the case that the laser best suited for reconstructing the mast motion by itself is the one that at this time appears less likely to fail.

\section{Discussion}
\label{sec:summary}

Through analysis of the mast behavior over time, as a function of SAA and of orbital phase, we have developed a method for successful PSF reconstruction should one of the metrology lasers on board \nustar\ fail. We do this by using the remaining PSD track, the SAA, the date, and the orbital phase to generate a synthetic mast file that can be used in the \nustar\ pipeline in order to produce sufficiently accurate data products. We found that, except for two narrow ranges of SAA, this method can produce reconstructed PSFs that are slightly larger but otherwise undistorted and, more importantly, exhibit no spectral changes. When compared with the PSF from the original two-laser solution, the central emission is slightly smeared out, but with an extraction region of 50\,arcseconds or greater, differences in the data products between the one- and two-laser scenarios are negligible. Therefore, we are satisfied that acceptable PSF reconstruction can be performed with a single metrology laser.

These results are a promising sign that the loss of a metrology laser on \nustar\ would not be a mission-ending problem. We plan to implement the generation of single-laser approximation mast files alongside the ordinary running of the \nustar\ pipeline, in order to confirm that these results hold up for future data and to easily enable a switch to using this method alone should a metrology laser fail within the next few years. 

The loss of a metrology laser would have implications for the scheduling of observations in order to maintain high-quality PSF reconstruction. The SAA ranges 74--78 and 114--116 would need to be avoided in order to prevent significant PSF distortion. Additionally, because we are unable to calibrate any of our SAA functions down to SAA = 0, this may have adverse effects on the usefulness of Solar data. In the case of LASER0 failing instead of LASER1, the SAA ranges to avoid would be larger, and the performance of PSF reconstruction generally worse. However, it may be possible to address the issue of bad SAA ranges by finding more sophisticated ways of modeling or adjusting the single-laser {\tt mast} file to more closely match the two-laser version, such as by investigating ways by which the shape of the $T_x$ and $T_y$ variations can be modeled rather than approximated by the measured laser track. We are also investigating an alternative machine-learning approach to the problem which may be able to successfully reconstruct the PSF even in SAA ranges that are problematic using this method.

Looking forward, there are positive implications for future missions seeking to use deployable masts in order to achieve a long focal length, since we have demonstrated that, as long as sufficient records of SAA and orbital phase are kept over the duration of the mission, it is possible to create a solid contingency plan for if a metrology system fails.

\subsection*{Acknowledgments}
We thank our anonymous reviewers for helpful comments that improved this paper. This work was supported under NASA Contract No. NNG08FD60C and made use of data from the \nustar\ mission, a project led by the California Institute of Technology, managed by the Jet Propulsion Laboratory, and funded by the National Aeronautics and Space Administration. We thank the \nustar\ Operations, Software, and Calibration teams for support with the execution and analysis of these observations. This research has made use of the \nustar\ Data Analysis Software (NuSTARDAS) jointly developed by the ASI Science Data Center (ASDC, Italy) and the California Institute of Technology.

%%%%% References %%%%%

\bibliography{laserspaper}   % bibliography data

\begin{thebibliography}{1}

\bibitem{harrison13}
F.~A. {Harrison}, W.~W. {Craig}, F.~E. {Christensen}, {\em et~al.}, ``{The
  Nuclear Spectroscopic Telescope Array (NuSTAR) High-energy X-Ray Mission},''
  {\em \apj} {\bf 770}, 103  (2013).

\bibitem{liebe10}
C.~C. {Liebe}, J.~{Burnham}, R.~{Cook}, {\em et~al.}, ``Metrology system for
  measuring mast motions on the nustar mission,'' in {\em 2010 IEEE Aerospace
  Conference},  1--11  (2010).

\bibitem{harp10}
D.~I. {Harp}, C.~C. {Liebe}, W.~{Craig}, {\em et~al.}, ``{NuSTAR: system
  engineering and modeling challenges in pointing reconstruction for a
  deployable x-ray telescope},'' in {\em Modeling, Systems Engineering, and
  Project Management for Astronomy IV},  G.~Z. {Angeli} and P.~{Dierickx},
  Eds., {\em Society of Photo-Optical Instrumentation Engineers (SPIE)
  Conference Series} {\bf 7738}, 77380Z  (2010).

\bibitem{liebe12}
C.~C. {Liebe}, B.~W. {Bauman}, G.~R. {Clark}, {\em et~al.}, ``{Design,
  Qualification, Calibration and Alignment of Position Sensing Detector for the
  NuSTAR Space Mission},'' {\em IEEE Sensors Journal} {\bf 12}, 2006--2013
  (2012).

\bibitem{forster16}
K.~{Forster}, K.~K. {Madsen}, H.~{Miyasaka}, {\em et~al.}, ``{Getting NuSTAR on
  target: predicting mast motion},'' in {\em Observatory Operations:
  Strategies, Processes, and Systems VI},  A.~B. {Peck}, R.~L. {Seaman}, and
  C.~R. {Benn}, Eds., {\em Society of Photo-Optical Instrumentation Engineers
  (SPIE) Conference Series} {\bf 9910}, 99100Z  (2016).

\bibitem{madsen20}
K.~K. {Madsen}, B.~W. {Grefenstette}, S.~{Pike}, {\em et~al.}, ``{NuSTAR low
  energy effective area correction due to thermal blanket tear},'' {\em arXiv
  e-prints} , arXiv:2005.00569  (2020).

\bibitem{an14}
H.~{An}, K.~K. {Madsen}, N.~J. {Westergaard}, {\em et~al.}, ``{In-flight PSF
  calibration of the NuSTAR hard X-ray optics},'' in {\em Space Telescopes and
  Instrumentation 2014: Ultraviolet to Gamma Ray},  T.~{Takahashi}, J.-W.~A.
  {den Herder}, and M.~{Bautz}, Eds., {\em Society of Photo-Optical
  Instrumentation Engineers (SPIE) Conference Series} {\bf 9144}, 91441Q
  (2014).

\bibitem{wik14}
D.~R. {Wik}, A.~{Hornstrup}, S.~{Molendi}, {\em et~al.}, ``{NuSTAR Observations
  of the Bullet Cluster: Constraints on Inverse Compton Emission},'' {\em \apj}
  {\bf 792}, 48  (2014).

\bibitem{madsen15}
K.~K. {Madsen}, F.~A. {Harrison}, C.~B. {Markwardt}, {\em et~al.},
  ``{Calibration of the NuSTAR High-energy Focusing X-ray Telescope.},'' {\em
  \apjs} {\bf 220}, 8  (2015).

\end{thebibliography}
\bibliographystyle{spiejour}   % makes bibtex use spiejour.bst

%%%%% Biographies of authors %%%%%

\vspace{2ex}\noindent\textbf{Hannah P. Earnshaw} received their PhD in astronomy from Durham University, UK, in 2017, and currently works as a data and calibration scientist for NuSTAR at California Institute of Technology. Their research interests include ultraluminous X-ray sources and transient extragalactic X-ray binaries.

\vspace{2ex}\noindent\textbf{Kristin K. Madsen} received her PhD in astrophysics from the University of Copenhagen in 2007. She has more than 10 years of experience in X-ray optics and astrophysical research and is one of the principal instrument and mission scientists for NuSTAR. She currently works as a staff scientist at Goddard Space Flight Center. Her research interests include active galactic nuclei, X-ray binaries, pulsar wind nebulae, and hard X-ray optical design.

\vspace{2ex}\noindent\textbf{Karl Forster} is a graduate of the University of Leicester and the University of Hertfordshire, UK, and received his PhD in astronomy from Columbia University. His research interest includes high-energy emission from active galactic nuclei, and he has developed and managed science operations for NASA missions GALEX and NuSTAR.

\vspace{2ex}\noindent\textbf{Brian W. Grefenstette} is a research scientist at California Institute of Technology. He is one of the principal NuSTAR mission scientists, working on calibration of the focal plane detectors. His science interests include nuclear astrophysics, X-ray binaries, supernovae, and supernova remnants.

\vspace{2ex}\noindent\textbf{Murray Brightman} received his PhD in astrophysics from Imperial College London in 2010 and is now a Staff Scientist at the NuSTAR Science Operations Center at Caltech. His research interests are broadly in the X-ray emission from black holes and neutron stars.

\vspace{2ex}\noindent\textbf{Andreas Zoglauer} received his PhD in 2005 from the Technische Universit\"{a}t M\"{u}nchen, Germany, for the development of novel simulation and data analysis tools for the MEGA telescope. He currently works at the University of California at Berkeley, and his research interests include the development of new simulation and analysis techniques, and the optimization and performance prediction of future instruments.

\vspace{2ex}\noindent\textbf{Fiona A. Harrison} received her PhD in physics from the University of California, Berkeley, in 1993. She is the California Institute of Technology (Caltech) Benjamin M. Rosen Professor of Physics and the Kent and Joyce Kresa Leadership Chair of the Division of Physics, Mathematics, and Astronomy. Her research is focused on the study of energetic phenomena ranging from gamma-ray bursts and black holes on all mass scales to neutron stars and supernovae. She is the principal investigator for NuSTAR.

\listoffigures
\listoftables

\end{spacing}
\end{document}